\documentclass[]{aastex631}
\usepackage{amsmath}

\newcommand{\Msun}{\mbox{\,$\rm M_{\odot}$}}
\newcommand{\Lsun}{\mbox{\,$\rm L_{\odot}$}}

\newcommand{\Teff}{$T_{\rm eff}$}

\newcommand{\logg}{$\log g$}

\newcommand{\co}{$\textrm{[C/O]}$}
\newcommand{\cfe}{$\textrm{[C/Fe]}$}
\newcommand{\ofe}{$\textrm{[O/Fe]}$}
\newcommand{\nfe}{$\textrm{[N/Fe]}$}
\newcommand{\sfe}{$\textrm{[s/Fe]}$}
\newcommand{\zrfe}{$\textrm{[Zr/Fe]}$}

\newcommand{\lsfe}{$\textrm{[ls/Fe]}$}
\newcommand{\hsfe}{$\textrm{[hs/Fe]}$}
\newcommand{\hsls}{$\textrm{[hs/ls]}$}
\newcommand{\feh}{$\textrm{[Fe/H]}$}

\shorttitle{$s$-process diversity in Galactic post-AGB stars with $Gaia$\,EDR3}
\shortauthors{Kamath et al.}
\graphicspath{{./}{figures/}}
\DeclareUnicodeCharacter{2212}{-}

\begin{document}

\title{Luminosities and masses of single Galactic Post-Asymptotic Giant Branch (Post-AGB) stars with distances from $Gaia$\,EDR3: The revelation of an $s$-process diversity}

\correspondingauthor{Devika Kamath}
\email{devika.kamath@mq.edu.au}

\author[0000-0001-8299-3402]{Devika Kamath}
\affil{Department of Physics and Astronomy, Macquarie University, Sydney, NSW, Australia}
\affil{Astronomy, Astrophysics and Astrophotonics Research Centre, Macquarie University, Sydney, NSW, Australia}

\author[0000-0001-5158-9327]{Hans Van Winckel}
\affiliation{Instituut voor Sterrenkunde, K.U.Leuven, Celestijnenlaan 200D bus 2401, B-3001, Leuven, Belgium}

\author[0000-0002-5026-6400]{Paolo Ventura}
\affil{INAF, Observatory of Rome, Via Frascati 33, 00077 Monte Porzio Catone (RM), Italy}

\author[0000-0002-7453-2945]{Maksym Mohorian}
\affil{Department of Physics and Astronomy, Macquarie University, Sydney, NSW, Australia}
\affil{Astronomy, Astrophysics and Astrophotonics Research Centre, Macquarie University, Sydney, NSW, Australia}

\author[0000-0003-1471-8892]{Bruce.J.Hrivnak}
\affiliation{Department of Physics and Astronomy, Valparaiso University, Valparaiso, IN 46383, USA}

\author[0000-0003-2442-6981]{Flavia Dell'Agli}
\affil{INAF, Observatory of Rome, Via Frascati 33, 00077 Monte Porzio Catone (RM), Italy}

\author[0000-0002-3625-6951]{Amanda Karakas}
\affiliation{School of Physics \& Astronomy, Monash University, Clayton VIC 3800, Australia}
\affiliation{ARC Centre of Excellence for All Sky Astrophysics in 3 Dimensions (ASTRO 3D)}




\begin{abstract}

Post-AGB stars are exquisite probes of AGB nucleosynthesis. However, the previous lack of accurate distances jeopardised comparison with theoretical AGB models. The $Gaia$ Early Data Release 3 ($Gaia$\,EDR3) has now allowed for a breakthrough in this research landscape. In this study, we focus on a sample of single Galactic post-AGBs for which chemical abundance studies were completed. We combined photometry with geometric distances to carry out a spectral energy distribution (SED) analysis and derive accurate luminosities. We subsequently determined their positions on the HR-diagram and compared this with theoretical post-AGB evolutionary tracks. While most objects are in the post-AGB phase of evolution, we found a subset of low-luminosity objects that are likely to be in the post-horizontal branch phase of evolution, similar to AGB-manqu\'e objects found in globular clusters. Additionally, we also investigated the observed bi-modality in the $s$-process enrichment of Galactic post-AGB single stars of similar \Teff\, and metallicities. This bi-modality was expected to be a direct consequence of luminosity with the $s$-process rich objects having evolved further on the AGB. However, we find that the two populations: the $s$-process enriched and non-enriched, have similar luminosities (and hence initial masses), revealing an intriguing chemical diversity. For a given initial mass and metallicity, AGB nucleosynthesis appears inhomogeneous and sensitive to other factors which could be mass-loss, along with convective and non-convective mixing mechanisms. Modelling individual objects in detail will be needed to investigate which parameters and processes dominate the photospheric chemical enrichment in these stars.

\end{abstract}

\keywords{stars: AGB and post-AGB --- stars: abundances --- Galaxy: abundances --- stars: evolution --- parallaxes}


\section{Introduction} 
\label{sec:intro}

Low- and intermediate-mass (LIM) stars (1\,$-$8\,\Msun) stars in their Asymptotic Giant Branch (AGB) phase of evolution are estimated to produce $\sim$\,90$\%$ of the solid material injected into the interstellar medium \citep{sloan08}, and are known to be one of the major producers of elements such as carbon, nitrogen and about half of the elements heavier than iron \citep{kobayashi20}. AGB stars are clearly key contributors to the chemical enrichment of the Universe. Traditionally, AGB stars are popularly used as tracers to quantify elemental isotopes \citep[e.g.,][]{garcia07a,hinkle16}. However, they pose many challenges for tracing AGB nucleosynthesis since their spectra are veiled by molecular lines \citep{abia08}, and modelling of their dynamical atmospheres is rather complex and uncertain \citep{perez-mesa19}.

In the evolution of LIM stars, post-Asymptotic Giant Branch (post-AGB) objects are considered to be those in transition between the AGB and the Planetary Nebula (PN) phase \citep[see][for a review]{vanwinckel03}. Post-AGB stars are typically of A\,$-$\,K spectral types, with a characteristic mid-IR excess, indicative of their dusty circumstellar environment. At the very end of the AGB phase, stars lose mass via a powerful wind, driven by stellar pulsations \citep{vw93} or by interaction with another star for stars in binary systems \citep{nie12, kamath16}. When the stellar envelope is reduced to a few hundredths of a solar mass, the post-AGB evolutionary phase starts \citep[e.g.][]{bertolami16}. Depending on the opacity of the circumstellar shell, the central star becomes exposed. Given their effective temperature, the photospheric spectra are dominated by atomic transitions, making post-AGB stars exquisite probes to examine the elements produced by the star during and prior to the AGB phase \citep[e.g.,][and references therein]{desmedt12,desmedt16}. Additionally, objects in the post-AGB phase (as shown in the Spectral Energy Distributions, SEDs, presented in Figure~\ref{sampleSED}) allow for full characterisation of the central star as well as the circumstellar material. 
The systematic identification and study of post-AGB stars began with the IRAS satellite $\sim$\,30 years ago \citep{kwok87}. Currently, the known sample of post-AGB stars are those residing in the Galaxy and the Magellanic Clouds (MCs). The Toru\'{n} catalogue \citep{szczerba07} lists around 391 likely optically-visible post-AGB candidates in the Galaxy, while systematic photometric and spectroscopic studies \citep{vanaarle11,kamath14,kamath15} have provided catalogues of  spectroscopically verified, optically visible post-AGB candidates in the Magellanic Clouds. 

Observational studies of the known sample of post-AGB stars in the Galaxy and the MCs have shown their SEDs can provide critical clues on the likely single or binary nature of these objects \citep[see][and references therein]{vanwinckel09,kamath15,oomen18}. Single post-AGBs (refereed to as 'shell-sources') show a distinct double-peaked SED, where the peak at optical wavelengths is representative of the central post-AGB star and the peak at infrared (IR) wavelengths is indicative of the detached dusty circumstellar envelope. The binary objects (referred to as 'disc-sources') show a broad onset in the near-IR, indicative of hot circumstellar dust that resides in a circumbinary disc around the post-AGB binary system \citep[see][and references therein]{vanwinckel17-pr}. 

The single or binary nature also has a significant effect on the observed photospheric abundances. As expected, many single post-AGB stars show signatures of AGB nucleosynthesis, such as enhancements of carbon (C), nitrogen (N), oxygen (O), and $s$-process elements\footnote{$s$-process elements are those created via the slow neutron-capture nucleosynthesis \citep[see][and references therein.]{gallino98,karakas14a}}\citep[e.g.,][]{desmedt12,desmedt16}. For post-AGB stars in binary systems, the circumbinary disc has shown to play a key role in altering the photospheric chemistry, resulting in a 'depleted' chemical pattern, wherein the stellar photosphere is depleted in refractory elements \citep[e.g.,][]{oomen18,kamath19}.

In addition to the above-mentioned chemical patterns, post-AGB stars also show intriguing chemical diversities \citep{vanwinckel03,kamath20}. For instance, the study by \citet{vanwinckel03-pr} reported a subset of single Galactic post-AGB stars that show no signatures of $s$-process enhancements (referred to as 'have-nots'). Also, in our previous study, \citep{kamath17} we reported a subset of luminous single post-AGB stars (one in the SMC and two in the Galaxy) that showed no traces of carbon enhancements nor of $s$-process elements, and are likely to have failed the third dredge-up. These chemical diversities directly point to the complex and poorly understood chemical mixing and nucleosynthesis that occurs during and prior to the AGB phase of evolution in these stars. 

In this study, we focus on the observed bi-modality in the enrichment of $s$-process elements, first reported by \citet{vanwinckel03-pr} almost two decades ago, based on detailed chemical abundance studies of 17 single post-AGB stars in the Galaxy. In \citet{vanwinckel03-pr} the two groups of objects were referred to as 'haves' - those with significant $s$-process enrichment and 'have-nots', i.e., those without traces of $s$-process enhancements. It was noted that both the 'haves' and 'have-nots' were of similar effective temperatures and metallicities (\feh). Whether the bi-modality reflected a true distribution due to intrinsic AGB nucleosynthesis (which is highly dependent on initial mass and metallicity) or whether it was due to an observational bias in the selection criteria of well studied post-AGB stars remained unsolved.

The challenges in solving this chemical bi-modality have been two-fold. Firstly, owing to the brevity of this phase \citep[$<$\,10,000 years depending on metallicity;][]{bertolami16}, post-AGB stars are relatively rare, and finding sufficient numbers of them is difficult. Secondly, element production depends upon the initial mass and composition of the star. So-far, poorly known distances to stars in our Galaxy stymied the determination of accurate luminosities and initial masses to these objects. 
In this study, we exploit the early release of $Gaia$\,EDR3 data, which has provided the opportunity to obtain distances and hence luminosities to the sample of known single post-AGB stars in our Galaxy. By comparing the position in the HR diagram with evolutionary tracks, we can estimate the initial masses and finally, we investigate the intriguing $s$-process enrichment bi-modality observed in the single post-AGB stars in the Galaxy.

\section{Sample of Galactic Post-AGB single stars} \label{sec:target}

For this study, we include all optically visible single post-AGB stars that were classified as 'haves' (hereafter referred to as: '$s$-process enriched') and 'have-nots' (hereafter referred to as: 'non $s$-process enriched') in \citet{vanwinckel03,vanwinckel03-pr}. Additionally, we also include more recently studied single post-AGB objects in the Galaxy for which a reliable chemical abundance study has been carried out. The final sample of objects (see Table~\ref{table1}) comprises of 18 $s$-process enriched objects and 13 non $s$-process enriched objects. For convenience, Table~\ref{table1} provides not only the object name but also a corresponding index with which we will refer to individual objects throughout the paper. We note that in Table~\ref{table1} we have provided the most commonly used name for the object. For completeness, we have complied all other names (e.g., IRAS names, HR names, HD names, SAO names, etc.) in Table~\ref{table2} (see Appendix~\ref{sec:chemabund}). Additionally, in Figure~\ref{fig:mwmap} we show the distribution of the sample set (as a function of initial metallicity) in the Galaxy. The objects are field stars and distributed along the Galactic disk. The relevant chemical abundances taken from the literature for each of the individual objects are listed in Table~\ref{table3} of Appendix~\ref{sec:chemabund}. 

We note that all the objects in our sample are likely either single stars or stars with undetermined wide orbits such that the binary interaction during evolution is limited and they likely evolve as single stars. All objects show the characteristic mid-IR excesses due to their circumstellar dust shells and expanding molecular envelopes (as mentioned in Section~\ref{sec:intro}). This is evidenced by their double-peaked SED (see Appendix~\ref{sec:SEDs}). A few objects without detected dust (e.g., Object 22: HD\,107369, Object 25: HR\,6144, Object 31: HR\,7671) are high-latitude supergiants, also recognised to be in their post-AGB phase of evolution \citep[e.g.,][]{vanwinckel97a,giridhar05,reyniers05}. Additionally, photometric monitoring of the majority of these objects have revealed low-amplitude pulsations, with periods of 25 to 165 days for those with spectral types of G, F, A and much shorter ones for the B stars \citep{hrivnak10,hrivnak15,hrivnak21}. 

The majority of the northern objects have also been a part of systematic long-term radial velocity monitoring programmes started in 2009 with the KU Leuven HERMES spectrograph \citep{raskin11} on the 1.2-m Mercator telescope at the Roque de los Muchachos observatory; and or by the long-term radial velocity monitoring programme at the Dominion Astrophysical Observatory in Victoria \citep{hrivnak17}. None of the objects have shown signatures of long-term variability in their radial velocity, thus confirming their likely single evolutionary nature. Therefore, we address the target sample as being all single stars.


\begin{figure}[htp]
\plotone{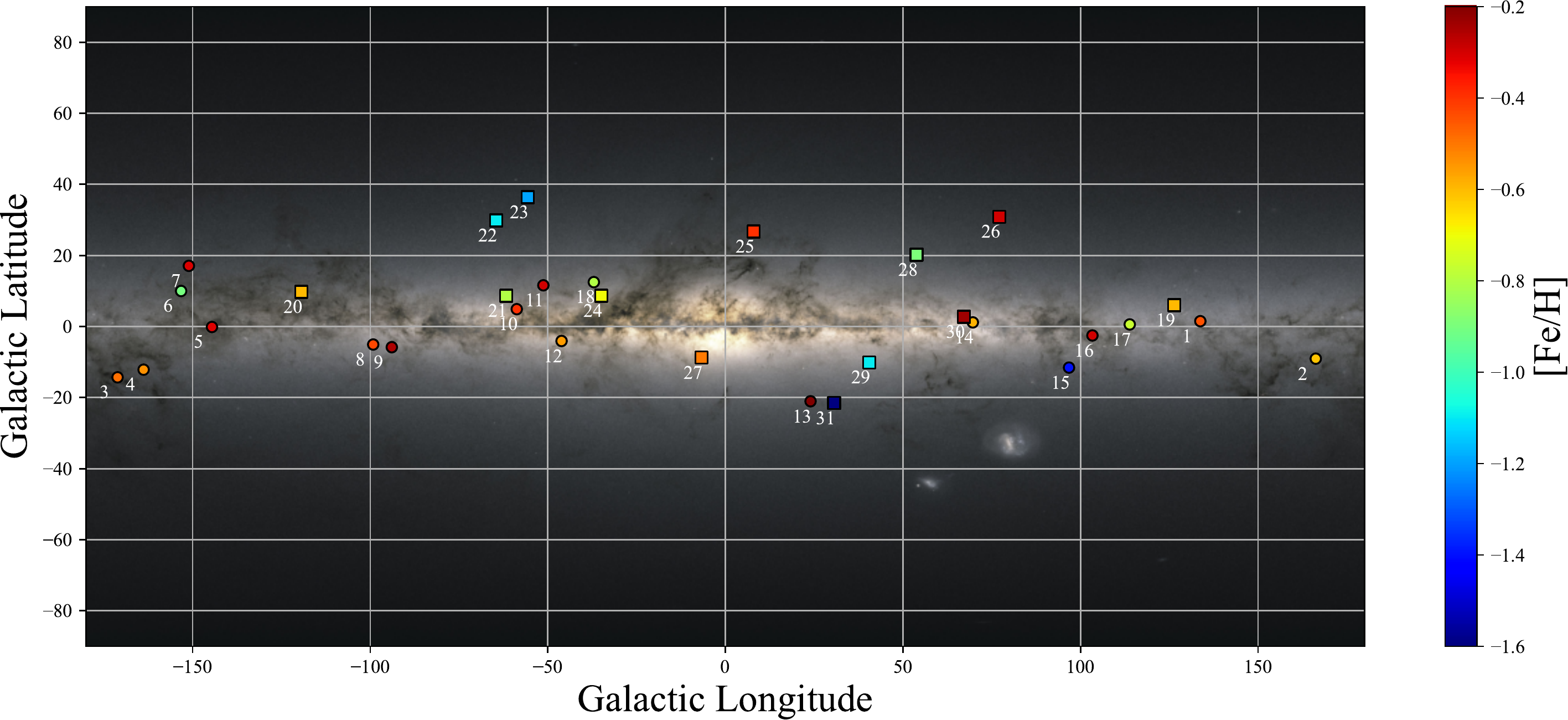}
\caption{An edge-on Milky Way map with the positions of the sample stars marked. The filled-circle symbols represent the $s$-process enriched rich sources. The filled-square symbols represent the non $s$-process enriched sources. The color bar represents the metallicity (\feh) of the objects. \label{fig:mwmap}}
\end{figure}

\section{Data Analysis: total reddening and photospheric luminosity estimation} \label{sec:lum+mass}

We use the latest available parallax data from the $Gaia$ Early Data Release 3 ($Gaia$\,EDR3). We present the $Gaia$\,EDR3 identifiers and coordinates in Table~\ref{table2} (see Appendix~\ref{sec:chemabund}). For the target sample, we obtained $Gaia$\,EDR3 parallaxes and their uncertainties, given with mas (milliarcseconds) accuracy (see Table~\ref{table1}). Table~\ref{table1} also lists the distance obtained by inverting the parallax (i.e., $z_{\rm (1/parallax)}$). However, it has been widely accepted that reliable distances cannot be obtained by inverting the parallax. To circumvent this issue, we use the distances from the study by \cite{BJ21}, in which the authors adopt a probabilistic approach to estimating stellar distances. The approach involves the use of a prior constructed from a three-dimensional model of our Galaxy which includes interstellar extinction and Gaia's variable magnitude limit. In particular, we adopt the Bailer-Jones geometric distances (i.e., $z_{\rm BJ}$, see Table~\ref{table1}). The geometric distances are derived from  the $Gaia$\,EDR3 parallaxes with a direction-dependent prior on distance. We note that \cite{BJ21} also provides photogeometric distances which used the color and apparent magnitude of the individual stars by including assumptions that stars of a given color have a restricted range of probable absolute magnitudes (plus extinction). However, this assumption does not hold for post-AGB stars. We therefore use the Bailer-Jones geometric distances (i.e., $z_{\rm BJ}$) and the associated upper and lower limits ($z_{\rm BJU}$, and $z_{\rm BJL}$, respectively) in our further analysis. These values are tabulated in Table~\ref{table1}.

Based on the reliability of the $Gaia$\,EDR3 astrometric parameters (and hence the corresponding Bailer-Jones geometric distances, $z_{\rm BJ}$), we categorise our target sample into two groups: Q1 - high quality and Q2 - low quality. We establish the quality flag (Q1 or Q2) based on the $Gaia$\,EDR3 renormalised unit weight error parameter \citep[RUWE, see][]{lindegren21}, wherein a RUWE\,$\gtrsim$\,1.4 indicates likely unreliable astrometry. The RUWE parameters and the quality flags for the individual sources are listed in Table~\ref{table1}. Objects with high RUWE objects could be partially resolved binary stars or tight astrometric binaries with a significant orbit-induced displacement of the photocenter \citep{lindegren21}. However, this does not apply to our target sample, since we address our target sample as being likely single stars or stars on very wide orbits and likely to evolve as single stars (see Section~\ref{sec:target}). Additionally, post-AGB objects can have a resolved nebula \citep[e.g.,][]{siodmiak08,lagadec11,ramoslarios12} which will impact the astrometric solution resulting in a higher RUWE value. Therefore, while we consider both the Q1 and Q2 objects in our data analysis, we only consider the Q1 objects while interpreting results and drawing conclusions (see Section~\ref{sec:DC}).

\begin{figure}[htp]
\centering
\includegraphics[width=9 cm]{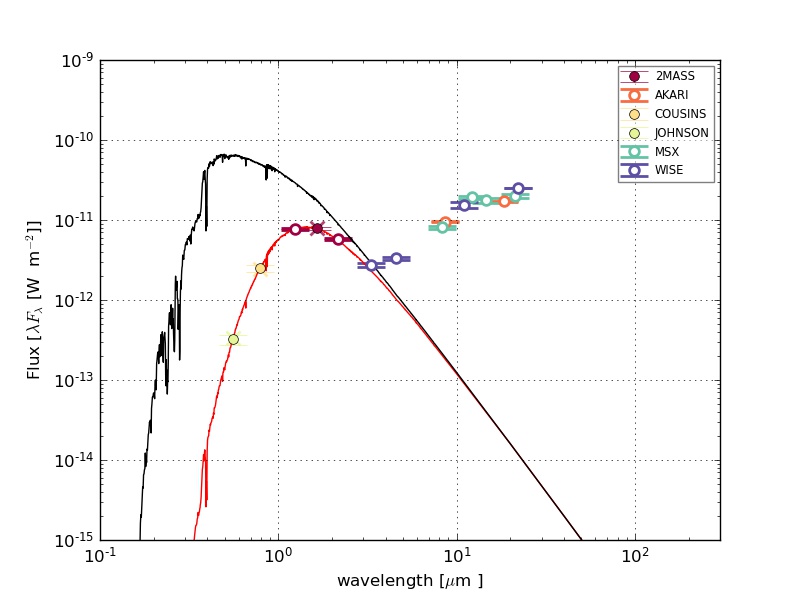}
\caption{SED of the $s$-process enriched post-AGB object, IRASZ02229+6208 (see Table~\ref{table1}). The colored symbols represent the photometry of different passband filters, with the appropriate photometric catalogue mentioned in the legend of the plot. The black solid line denotes the appropriate unreddened atmospheric model.The red line is the reddened model with E($B$-$V$)\,=\,1.9, fitted to the observed data. See text for more details. \label{sampleSED}}
\end{figure}

To estimate the luminosity of the central star, i.e., the photospheric luminosity, we need to individually derive the total reddening, E($B-V$), which includes interstellar and circumstellar reddening, for all the sources. We do this by systematically constructing SEDs for all objects in our sample and minimizing the difference between the optical and near-IR fluxes and the reddened atmospheric models. The photometric data points for all objects are automatically retrieved from the Vizier database \citep{ochsenbein00}. For the atmospheric models, we used the appropriate Kurucz atmospheric models \citep{castelli03}; the parameters of which were found in the spectroscopic analyses presented in previous studies (see last column of Table~\ref{table3}). For the SED fitting, we interpolate in the chi-square ($\chi^2$) region between the models centered around their spectroscopically determined parameters while also taking into account the uncertainty in the derived stellar parameters. 

Figure~\ref{sampleSED} shows the example of one of the SEDs and the results from the fitting procedure. The SED fits for the full sample of objects are presented in Appendix~\ref{sec:SEDs}. We note that we assume that the total extinction in the line-of-sight has the wavelength dependency of the interstellar-medium (ISM) extinction law \citet{cardelli89} with a R$_{\rm v}$\,$=$\,3.1. It is probable that the circumstellar extinction law is different from the interstellar extinction law, but this exploration is beyond the scope of this study. 

The uncertainty on the estimated total reddening parameter, E($B-V$), was computed by determining the confidence intervals of the free parameters. Subsequently, we determined the photospheric luminosity by integrating over all wavelengths the model atmosphere scaled which we scaled to the dereddened photometric data. We note that stellar variability is not taken into account and for high-amplitude variables, this translates into a higher $\chi^2$ value. For objects with a high reddening value, the uncertainty on the E($B-V$) can be significant (see Table~\ref{table1}). However, for the Q1 sources, the uncertainties on the E($B-V$) values are small (see Table~\ref{table1}), and hence the uncertainties on the luminosity are dominated by the uncertainties on the distances.

The derived luminosity (L/\Lsun) together with the upper and lower limits (L$_{\rm Upper}$/\Lsun\, and L$_{\rm Lower}$/\Lsun), as well as the estimated E($B-V$), \Teff, and \logg\, of the best fitting atmosphere model are tabulated in Table~\ref{table1}. L$_{\rm Upper}$/\Lsun\, and L$_{\rm Lower}$/\Lsun represent the impact of the uncertain distance (defined by $z_{\rm BJU}$, and $z_{\rm BJL}$) on the luminosity. While Table~\ref{table1} lists the derived luminosities for all the objects in our target sample, we remind the reader that only the Q1 objects are with reliable astrometry and hence with reliable distances and derived luminosities. We anticipate that the upcoming $Gaia$\,DR3 data release (planned for June 2022) will provide more reliable astrometry and the possibility to confirm the luminosities for the Q2 sources.

\begin{longrotatetable}
\begin{deluxetable*}{rrrrrrrrrrrrrrrr}
\tabletypesize{\footnotesize}
\tablecolumns{16}
\tablewidth{0pt}
\tabletypesize{\scriptsize}
\setlength{\tabcolsep}{2pt}
\tablecaption{Fundamental properties of the $s$-process rich and non-enriched single Galactic post-AGB stars. See Section~\ref{sec:lum+mass} for more details. \label{table1}}
\tablehead{
\colhead{Index} & \colhead{Object} & \colhead{Parallax} & \colhead{Error} & \colhead{RUWE} & \colhead{$z_{\rm (1/parallax)}$} & \colhead{$z_{\rm BJ}$} & \colhead{$z_{\rm BJL}$} & \colhead{$z_{\rm BJU}$} & \colhead{\Teff} & \colhead{\logg} &\colhead{E($B-V$)} & \colhead{L/\Lsun} & \colhead{L$_{\rm Lower}$/\Lsun} & \colhead{L$_{\rm Upper}$/\Lsun} & \colhead{Flag} \\
\colhead{} & \colhead{} & \colhead{(mas)} & \colhead{(mas)} &  \colhead{} & \colhead{(pc)} & \colhead{(pc)} & \colhead{(pc)} & \colhead{(pc)} & \colhead{(K)} & \colhead{(dex)} & \colhead{} & \colhead{} & \colhead{} & \colhead{} & \colhead{}\\
%
} 
\startdata
\multicolumn{16}{c}{Post-AGB stars with $s$-process enrichment}\\
\hline
1	&	IRAS\,Z02229+6208	&	0.38	&	0.06	&	2.5	&	2627.53	&	2352.18	&	2063.48	&	2687.03	&	5952 $\pm$ 250	&	0.00	&	1.90$^{+0.08}_{-0.42}$	&	12959	&	9973	&	16911	&	Q2	\\
2	&	IRAS\,04296+3429	&	-0.38	&	0.17	&	5.8	&	-2635.11	&	5048.41	&	3899.38	&	7150.92	&	7272 $\pm$ 250	&	0.73	&	2.03$^{+0.06}_{-0.19}$	&	10009	&	5971	&	20082	&	Q2	\\
3	&	IRAS\,05113+1347	&	-0.01	&	0.15	&	6.7	&	-108371.7	&	4629.82	&	3312.65	&	8416.2	&	5025 $\pm$ 250	&	0.01	&	0.75$^{+0.35}_{-0.09}$	&	2037	&	1043	&	6731	&	Q2	\\
4	&	IRAS\,05341+0852	&	0.51	&	0.19	&	13.0	&	1960.07	&	2057.38	&	1603.89	&	2778.95	&	6274 $\pm$ 250	&	0.84	&	1.18$^{+0.19}_{-0.06}$	&	324	&	197	&	592	&	Q2	\\
5	&	IRAS\,06530-0213	&	0.24	&	0.07	&	3.7	&	4145.93	&	3777.67	&	2886.46	&	4990.04	&	7809 $\pm$ 250	&	1.70	&	1.85$^{+0.02}_{-0.18}$	&	4687	&	2736	&	8178	&	Q2	\\
6	&	IRAS\,07134+1005	&	0.45	&	0.02	&	0.9	&	2203.76	&	2099.09	&	1991.41	&	2209.25	&	7485 $\pm$ 250	&	0.50	&	0.43$^{+0.10}_{-0.22}$	&	5505	&	4955	&	6098	&	Q1	\\
7	&	IRAS\,07430+1115	&	3.06	&	0.5	&	21.8	&	327.04	&	360.93	&	299.59	&	442.65	&	5519 $\pm$ 250	&	1.43	&	1.04$^{+0.38}_{-0.12}$	&	20	&	14	&	30	&	Q2	\\
8	&	IRAS\,08143-4406	&	0.24	&	0.02	&	1.4	&	4198.54	&	4154.69	&	3877.12	&	4568.46	&	7013 $\pm$ 250	&	1.31	&	1.53$^{+0.95}_{-0.95}$	&	4509	&	3927	&	5452	&	Q1	\\
9	&	IRAS\,08281-4850	&	-0.14	&	0.07	&	6.0	&	-7300.68	&	11452.58	&	8728.27	&	15113.73	&	7462 $\pm$ 250	&	1.04	&	1.23$^{+0.11}_{-0.04}$	&	9584	&	5567	&	16692	&	Q2	\\
10	&	IRAS\,12360-5740	&	0.09	&	0.01	&	1.1	&	10970.73	&	9082.07	&	8261.41	&	10230.48	&	7273 $\pm$ 250	&	1.59	&	1.01$^{+0.35}_{-0.35}$	&	6258	&	5178	&	7940	&	Q1	\\
11	&	IRAS\,13245-5036	&	0.01	&	0.02	&	1.6	&	85919.8	&	14207.28	&	11305.73	&	17383.96	&	9037 $\pm$ 250	&	3.20	&	0.64$^{+0.14}_{-0.09}$	&	11221	&	7106	&	16800	&	Q2	\\
12	&	IRAS\,14325-6428	&	0.19	&	0.04	&	2.2	&	5220.46	&	4883.39	&	4261.63	&	5811.07	&	7256 $\pm$ 250	&	1.00	&	1.07$^{+0.22}_{-0.17}$	&	4935	&	3758	&	6988	&	Q2	\\
13	&	IRAS\,14429-4539	&	-0.11	&	0.51	&	2.8	&	-9372.15	&	3847.71	&	2160.1	&	6548.2	&	9579 $\pm$ 250	&	2.48	&	2.63$^{+0.36}_{-0.51}$	&	5049	&	1591	&	14624	&	Q2	\\
14	&	IRAS\,19500-1709	&	0.4	&	0.03	&	1.0	&	2504.9	&	2310.24	&	2164.96	&	2481.49	&	8239 $\pm$ 250	&	1.08	&	0.56$^{+0.03}_{-0.07}$	&	7053	&	6194	&	8138	&	Q1	\\
15	&	IRAS\,20000+3239	&	0.2	&	0.05	&	2.2	&	4880.07	&	4581.29	&	3695.53	&	6075.01	&	5478 $\pm$ 250	&	0.13	&	1.76$^{+0.09}_{-0.46}$	&	14342	&	9332	&	25218	&	Q2	\\
16	&	IRAS\,22223+4327	&	0.33	&	0.03	&	1.7	&	3007.27	&	2678.03	&	2546.9	&	2878.56	&	6008 $\pm$ 250	&	1.05	&	0.43$^{+0.28}_{-0.06}$	&	2163	&	1956	&	2499	&	Q2	\\
17	&	IRAS\,22272+5435	&	0.69	&	0.03	&	1.2	&	1457.75	&	1409.84	&	1355.87	&	1464.67	&	5325 $\pm$ 250	&	0.77	&	0.88$^{+0.34}_{-0.08}$	&	5659	&	5234	&	6108	&	Q1	\\
18	&	IRAS\,23304+6147	&	0.24	&	0.03	&	1.6	&	4226.42	&	3979.67	&	3620.05	&	4390.37	&	6276 $\pm$ 250	&	0.78	&	1.83$^{+0.17}_{-0.20}$	&	7712	&	6381	&	9386	&	Q2	\\
\hline
\multicolumn{16}{c}{Post-AGB stars without $s$-process enrichment}\\ 
\hline
19	&	IRAS\,01259+6823	&	0.62	&	0.14	&	1.3	&	1624.61	&	1781.38	&	1434.96	&	2456.33	&	5510 $\pm$ 250	&	2.50	&	1.02$^{+0.24}_{-0.07}$	&	340	&	220	&	646	&	Q1	\\
20	&	IRAS\,08187-1905	&	0.29	&	0.03	&	1.7	&	3473.16	&	3258.96	&	2917.24	&	3649.91	&	5772 $\pm$ 250	&	0.98	&	0.07$^{+0.31}_{-0.02}$	&	2619	&	2099	&	3286	&	Q2	\\
21	&	SAO\,239853	&	-0.01	&	0.07	&	3.7	&	-117255.41	&	8691.08	&	6485.52	&	12490.93	&	7452 $\pm$ 250	&	1.49	&	0.30$^{+0.08}_{-0.08}$	&	23490	&	13080	&	48520	&	Q2	\\
22	&	HD\,107369	&	0.37	&	0.02	&	1.1	&	2725.26	&	2568.38	&	2429.15	&	2705.92	&	7533 $\pm$ 250	&	2.45	&	0.07$^{+0.13}_{-0.05}$	&	910	&	814	&	1010	&	Q1	\\
23	&	HD\,112374	&	0.57	&	0.02	&	1.0	&	1763.78	&	1684.48	&	1619.4	&	1768.68	&	6393 $\pm$ 250	&	0.80	&	0.30$^{+0.10}_{-0.28}$	&	10777	&	9961	&	11882	&	Q1	\\
24	&	HD\,133656	&	0.56	&	0.03	&	0.9	&	1776.54	&	1707.63	&	1646.84	&	1781.63	&	8238 $\pm$ 250	&	1.38	&	0.29$^{+0.01}_{-0.08}$	&	5227	&	4861	&	5690	&	Q1	\\
25	&	HR\,6144	&	0.28	&	0.03	&	1.2	&	3561.19	&	3101.16	&	2894.87	&	3387.69	&	6728 $\pm$ 250	&	0.93	&	0.11$^{+0.15}_{-0.01}$	&	25491	&	22212	&	30419	&	Q1	\\
26	&	HD\,161796	&	0.5	&	0.02	&	1.2	&	1991.19	&	1920.96	&	1829.56	&	2015.66	&	6139 $\pm$ 250	&	0.99	&	0.13$^{+0.45}_{-0.13}$	&	5742	&	5209	&	6322	&	Q1	\\
27	&	IRAS\,18025-3906	&	0.54	&	0.19	&	8.6	&	1865.62	&	3046.67	&	1973.6	&	7194.98	&	6154 $\pm$ 250	&	1.18	&	0.96$^{+0.35}_{-0.17}$	&	2324	&	975	&	12963	&	Q2	\\
28	&	HD\,335675	&	0.03	&	0.18	&	13.7	&	30507.45	&	4888.53	&	3319.14	&	6540.55	&	6082 $\pm$ 250	&	1.58	&	0.85$^{+0.20}_{-0.04}$	&	15843	&	7303	&	28359	&	Q2	\\
29	&	IRAS\,19386+0155	&	0.32	&	0.16	&	11.6	&	3088.79	&	3631.35	&	2441.76	&	5588.81	&	6303 $\pm$ 250	&	1.00	&	1.23$^{+0.35}_{-0.14}$	&	9611	&	4345	&	22765	&	Q2	\\
30	&	IRAS\,19475+3119	&	0.32	&	0.02	&	1.4	&	3165.15	&	2971.43	&	2785.82	&	3135.57	&	8216 $\pm$ 250	&	1.01	&	0.61$^{+0.04}_{-0.16}$	&	6775	&	5955	&	7545	&	Q1	\\
31	&	HR\,7671	&	1.34	&	0.03	&	0.8	&	748.77	&	727.4	&	714.04	&	742.99	&	6985 $\pm$ 250	&	0.83	&	0.40$^{+0.11}_{-0.18}$	&	3579	&	3449	&	3734	&	Q1	\\
\enddata
\end{deluxetable*}
\end{longrotatetable}

\section{Positions of the Galactic post-AGB single stars in the HR-diagram}
\label{sec:hr}
\subsection{Evolutionary nature}

\begin{figure*}
\gridline{\fig{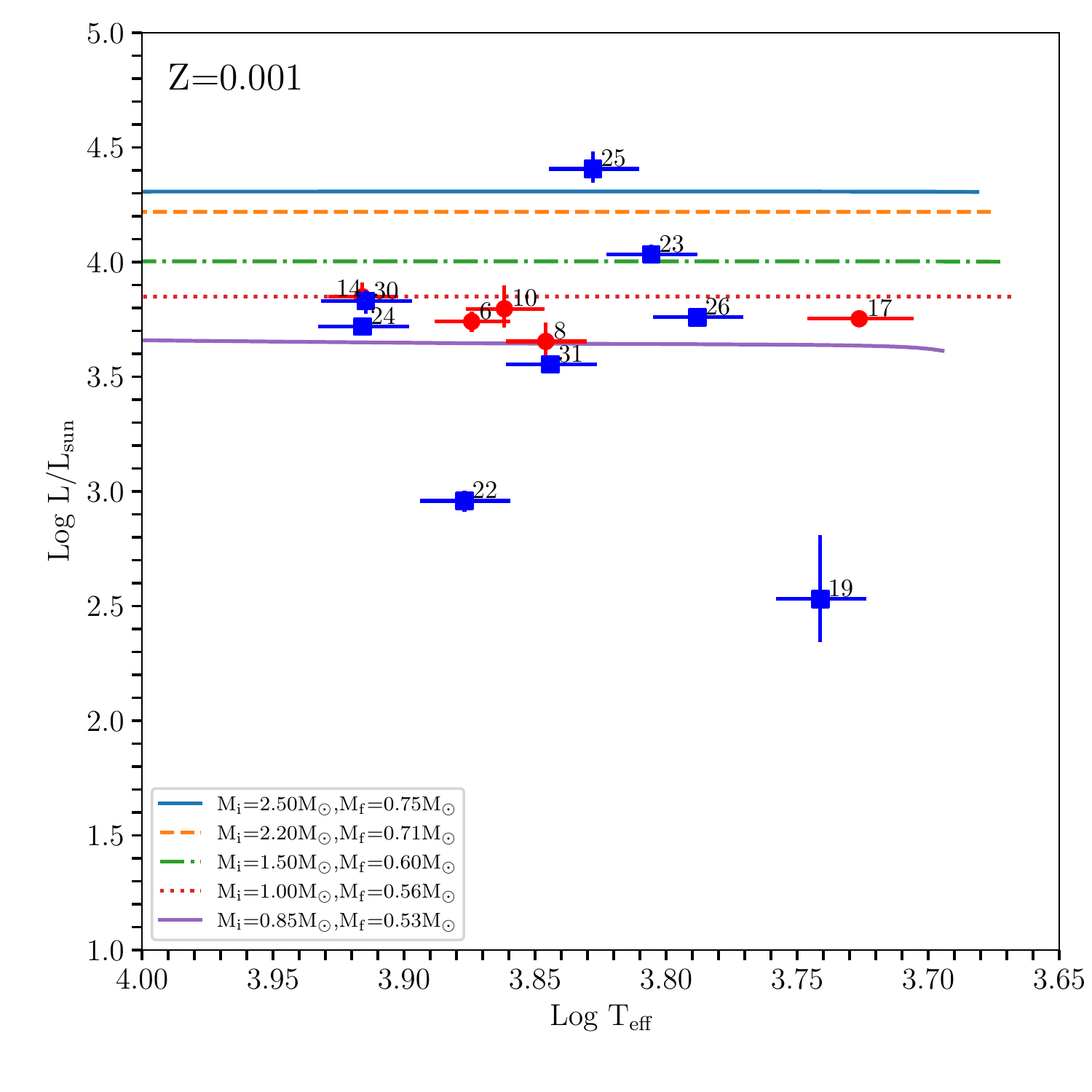}{0.5\textwidth}{}
          \fig{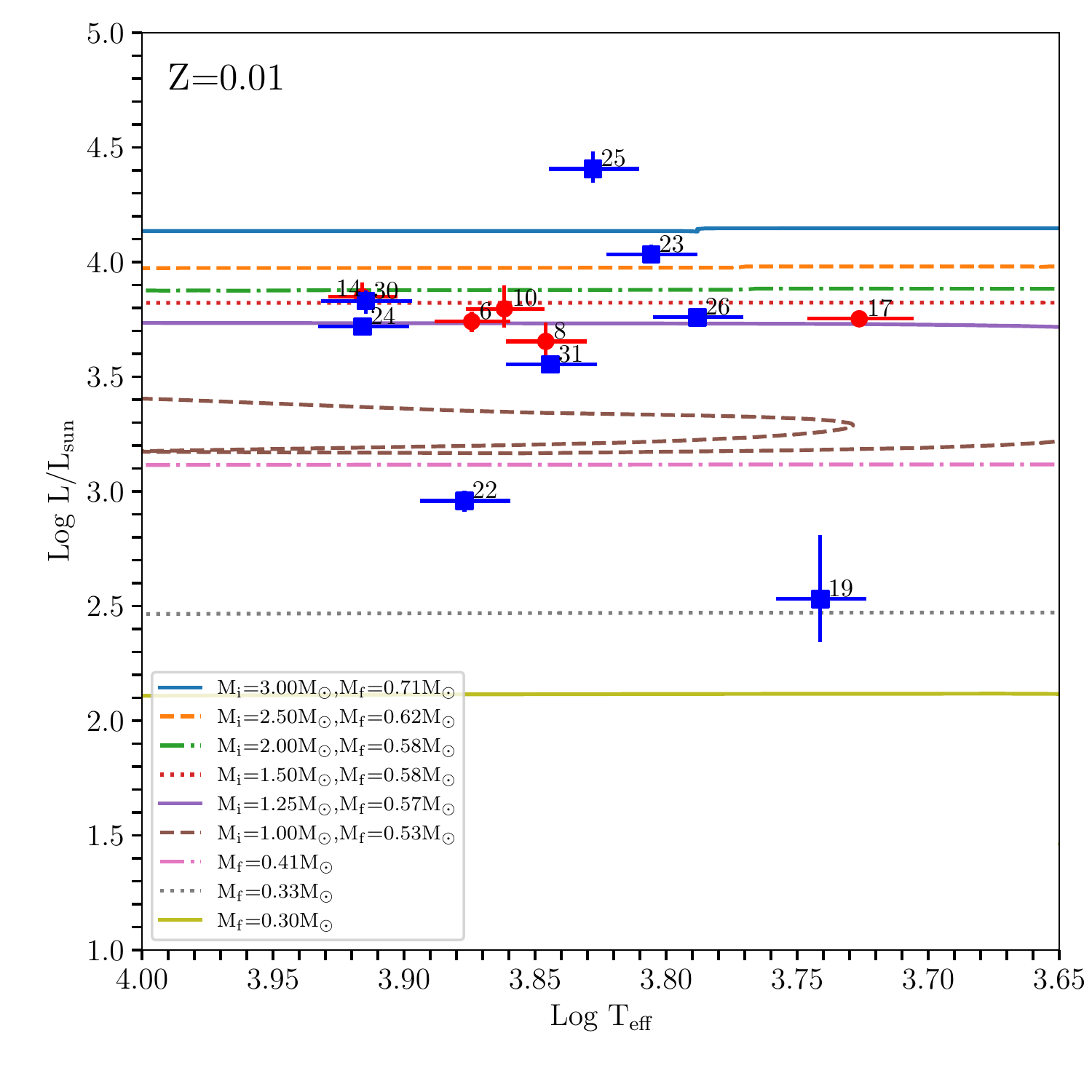}{0.5\textwidth}{}
          }
\caption{Positions of the Q1 post-AGB stars in the HR diagram. The red filled-circles represent the $s$-process enriched objects and the blue filled-squares represent the non $s$-process enriched objects. The numbers represent the individual object numbers as listed in Table~\ref{table1}. Also shown are the available post-AGB evolutionary tracks of \citet{bertolami16} with $Z$\,=\,0.001 (left panel) and $Z$\,=\,0.01 (right panel). See text for more details. \label{fig:hr}}
\end{figure*}

To better understand the evolutionary stage of the target post-AGB objects with reliable $Gaia$\,DR3 astrometry and hence luminosities, we investigate the positions of the Q1 objects in the HR-diagram (see Figure~\ref{fig:hr}). For the sake of completeness, we also present the HR-diagram with both Q1 and Q2 objects in Appendix~\ref{sec:hr_p1_p2}, but we do not further consider the Q2 objects.

In Figure~\ref{fig:hr}, the $s$-process enriched Q1 objects are plotted as red filled-circles, and the non $s$-process enriched Q1 objects are plotted as blue filled-squares. The effective temperatures (\Teff) and photospheric luminosities (L/\Lsun) are those derived from the SED fitting explained above. Also shown in Figure~\ref{fig:hr} are the available evolutionary tracks of \citet{bertolami16} for post-AGB stars covering the final masses in the range 0.83\Msun to 0.53\Msun. Since our target sample ranges from -1.5\,$\lesssim$\feh\,$\lesssim$-0.3 (see Table~\ref{table2}), we consider two different initial metallicities. The left panel of Figure~\ref{fig:hr} shows tracks with an initial metallicity of $Z$\,=\,0.001 (i.e., \feh\,$\approx$\,$-$1.5). The right panel of Figure~\ref{fig:hr} shows tracks with an initial metallicity of $Z$\,=\,0.01 (i.e., \feh\,$\approx$\,$-$0.30). Based on the positions of our Q1 targets on the HR-diagram, we find that the majority of the objects sit within the final mass range of 0.53\Msun to 0.83\Msun which translates to the initial mass range of $\sim$0.9\Msun\, to 3.00\Msun, based on the empirically derived initial masses from the post-AGB evolutionary tracks of \citet{bertolami16}. 

We find that two non $s$-process enriched stars (Object 19: IRAS\,01259+6823 and Object 22: HD\,107369) have luminosities $<$1000\Lsun, suggesting that they never reached the AGB phase. These luminosities are significantly smaller than the typical luminosity of the tip of the RGB (i.e., $\sim$\,2000\,$-$\,2500\,\Lsun) and are similar to those of the dusty post-RGB stars \citep{kamath16} which are the low-luminosity analogues of post-AGB objects. However, the dusty post-RGB stars are likely to be in binary systems, wherein the mass loss driven by the binary interaction pre-maturely terminated the RGB evolution, resulting in the objects evolving off the RGB as post-RGB objects. As explained in Section~\ref{sec:target} the post-AGB stars are likely to be single objects.


A possibility is that these stars have completed their Horizontal Branch (HB) evolution and are currently evolving through a post-HB phase that began after helium was exhausted in the core. The study by \citet{greggio90} showed that when the mass of the envelope above the helium core at the beginning of the HB phase is reduced to a few hundredths of a solar mass, the stars barely reach the AGB phase. Instead, after the HB evolution, and a relatively short expansion phase, they  start  contracting,  evolving  through the so called AGB-manqu\'e phase \citep{greggio90}. According to this interpretation such objects would be the counterparts of the stars populating the blue side of the HBs of some Galactic Globular Clusters, such as NGC\,2419 \citep{ripepi07} and NGC\,2808 \citep{bedin00}. They descend from low-mass ${\rm M}<1~{\rm M}_{\odot}$ progenitors, thus they are likely to be the oldest objects in the sample. 


Detailed evolutionary tracks of stars starting from the HB and extended until the start of the WD cooling track (for a range of masses between $\sim$0.5 to 0.6\Msun\, and at appropriate initial metallicities) will be required to test whether these low-luminosity stars are indeed in the AGB-manqu\'e phase. This is outside the scope of this observational paper, but will be pursued in Paper II. 



\subsection{Chemical diversity}

Interestingly, as shown in Figure~\ref{fig:hr}, we find that there is no obvious separation in the luminosities (and hence current and initial masses) of the majority of the $s$-process enriched stars and the non $s$-process enriched stars. This reveals that the observed bi-modality in the $s$-process enrichment (see Section~\ref{sec:intro}) is not purely a direct consequence of initial mass, as previously expected \citep{gallino98,karakas16}. In Section~\ref{sec:DC} we further explore the chemical composition trends of the two classes of objects. We note that the current/initial mass is highly dependent on metallicity. In our subsequent study (Kamath et al., in-prep; hereafter referred to as Paper II), we are calculating dedicated stellar models, tailored to the appropriate stellar parameters of the individual objects, which will provide accurate current and initial mass estimates and also allow for investigating the true nature of the individual objects based on their positions on the HR-diagram as well as their observed chemical abundances.

 \section{Discussion and Conclusions} \label{sec:DC}
 
 \begin{figure*}[htp]
\gridline{\fig{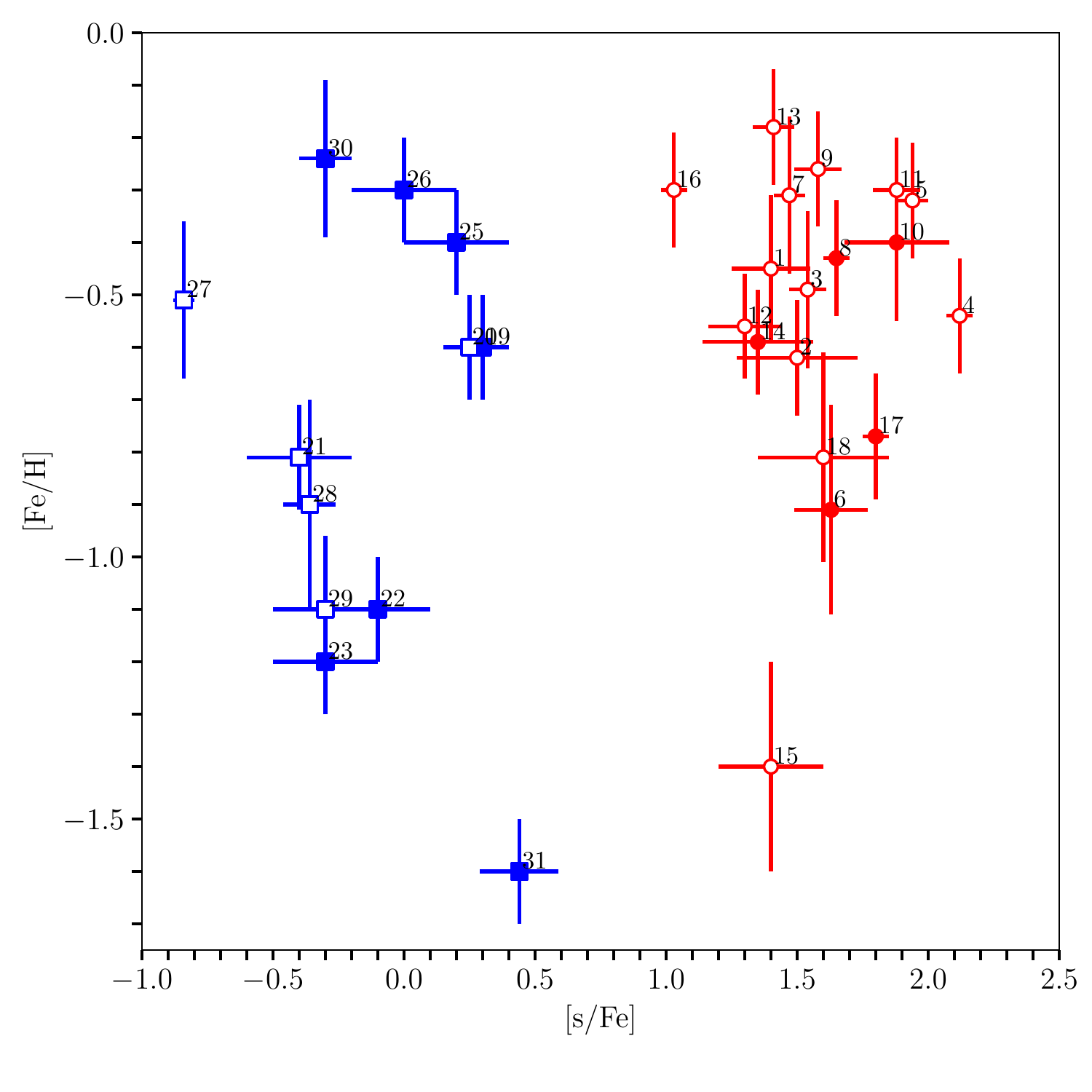}{0.5\textwidth}{}
          \fig{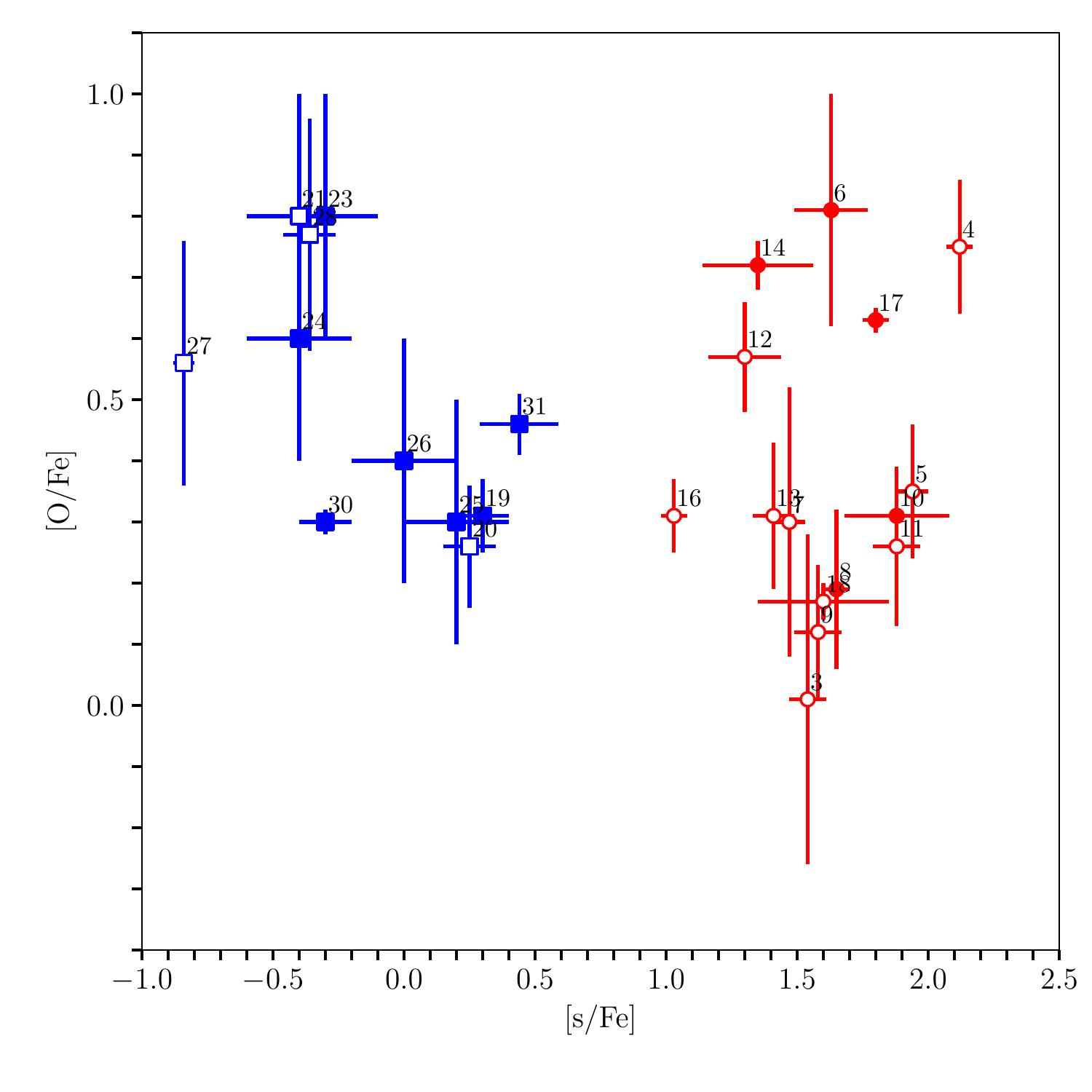}{0.5\textwidth}{}
          }
\vskip-40pt
\gridline{\fig{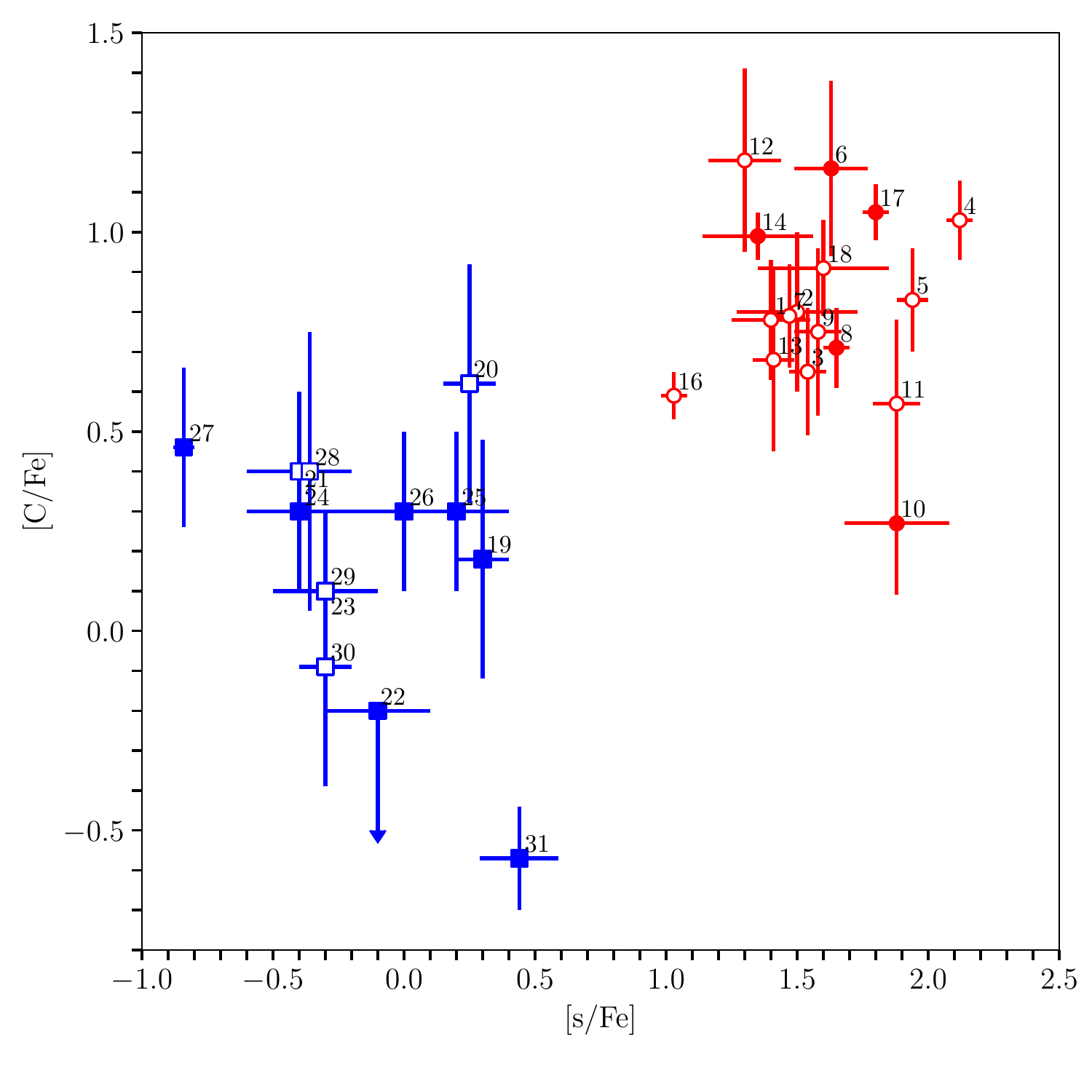}{0.5\textwidth}{}
          \fig{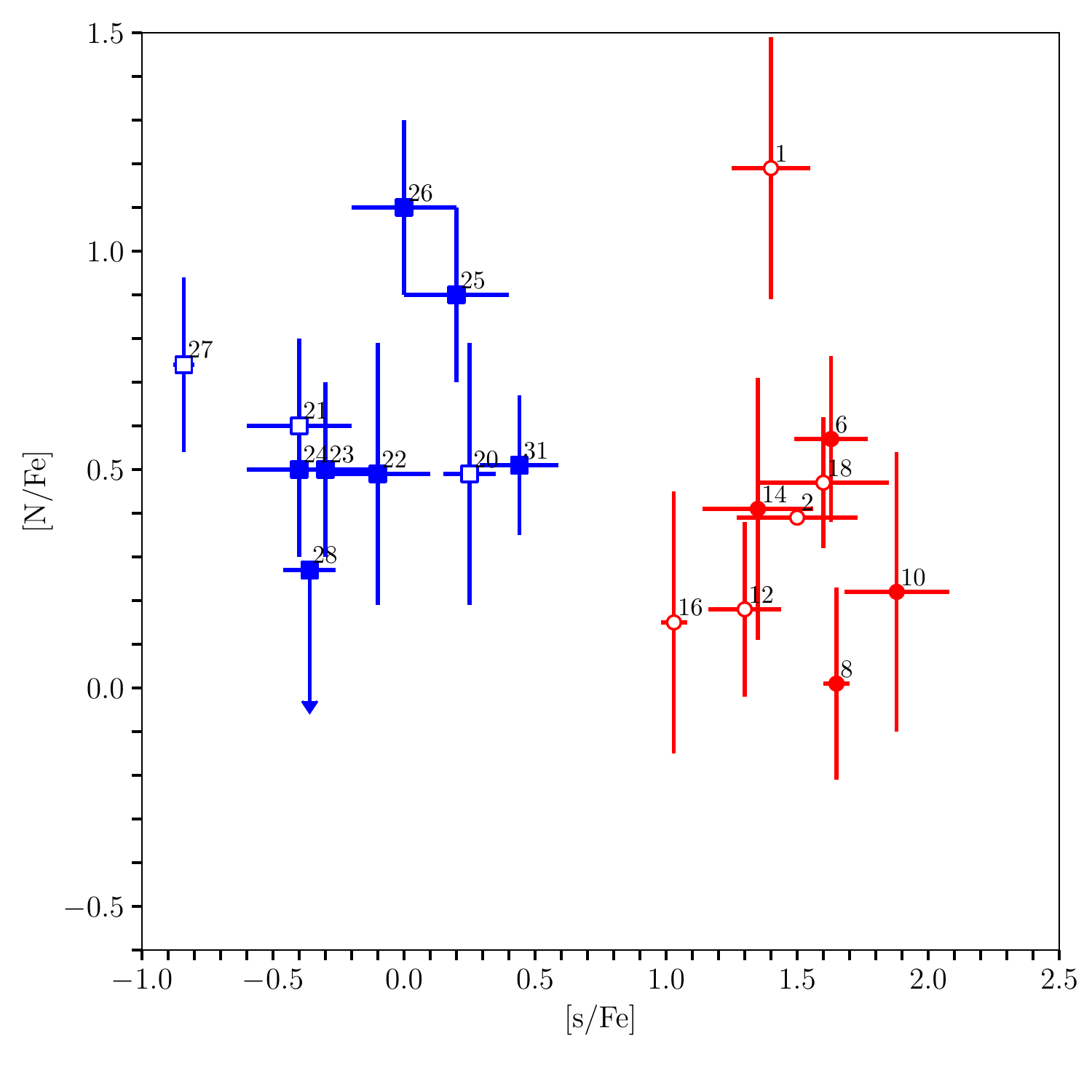}{0.5\textwidth}{}
          }
\caption{The \feh (top-left), \ofe (top-right), \cfe (bottom-left), and \nfe (bottom-right) ratios as a function of \sfe\, for the sample of post-AGB stars. The red circles represent the $s$-process enriched objects and the blue squares represent the non $s$-process enriched objects. Filled symbols denote the Q1 objects and the open symbols denote the Q2 objects. The numbers represent the individual object numbers as listed in Table~\ref{table1}. The derived abundances and errors are presented in Table~\ref{table3}. See Section~\ref{sec:DC} for more details.}
\label{fig:CNO}
\end{figure*}

The derived luminosities to the Galactic Q1 post-AGB stars in our target sample together with their detailed surface compositions has allowed us to confirm the evolutionary nature of these objects (via their positions in the HR diagram, see Figure~\ref{fig:hr}) and draw important conclusions on the nucleosynthesis that occurred during and, to some extant, prior to the AGB phase.

The positions of the $s$-process enriched and non $s$-process enriched stars in the HR diagram (Figure~\ref{fig:hr}) has firstly confirmed that the majority of the objects in the target sample are indeed residing in the post-AGB phase of their lives. A small subset of objects (see Section~\ref{sec:hr}) are likely to have had their evolution cut short and are likely to reside on post-HB phase. 

Additionally, as mentioned in Section~\ref{sec:intro}, it was previously believed that the long-standing expectation that the $s$-process bi-modality observed in single Galactic post-AGB stars is due to differences in their luminosities (and hence initial masses), with the $s$-process enriched expected to be more luminous than the non $s$-process enriched population \citep{vanwinckel03-pr}. Our results show that there is no obvious distinction between the current luminosities (and hence current- and initial-masses) of the two populations.

To further investigate the two populations, we consider their metallicities and their derived abundance ratios of the CNO elements as a function of their $s$-process enrichment (see Table~\ref{table2} in Appendix~\ref{sec:chemabund}). In Figure~\ref{fig:CNO}, we present the \feh, \ofe, \cfe, and \nfe\,  ratios' as a function of \sfe\, for the target sample of post-AGB objects (see Figure~\ref{fig:CNO}). We maintain the same color-coding and symbols as before to represent the $s$-process enriched stars and the non $s$-process enriched stars. 

The top-left panel of Figure~\ref{fig:CNO} which shows \feh\, versus\, \sfe\, demonstrates that there is no obvious effect of \feh\, on the $s$-process bi-modality. The objects from both groups show a spread in metallicity ranging from -1.5\,$\lesssim$\feh\,$\lesssim$-0.3. A similar conclusion can be drawn from the plot of the \ofe\, versus\, \sfe\, (top-right panel of Figure~\ref{fig:CNO}). 

On the other hand, the plot showing the \cfe\, versus\, \sfe\, (bottom-left panel of Figure~\ref{fig:CNO}) shows that, as expected from standard AGB nucleosynthesis theories \citep[e.g.,][]{fishlock14, ventura18, karakas18}, the majority of the $s$-process enriched population are more carbon-enhanced (with \cfe\,$\gtrsim$\,0.5) while the non $s$-process enriched population have \cfe\,$\lesssim$\,0.5. 

The \nfe\, versus\, \sfe\, plot (bottom-right panel of Figure~\ref{fig:CNO}), which shows all the objects with available \nfe\, ratios, reveals that for the $s$-process enriched population, the nitrogen enhancement is on average slightly lower compared to the non $s$-process enriched population. By and large, based on Figures~\ref{fig:hr} and \ref{fig:CNO}, it can be concluded that the population of stars showing $s$-process enrichment as well as the ones not showing $s$-process enhancements, are mostly within the initial-mass range of 1\,to\,3\Msun\, and with sub-solar metallicity.

We note that reliable astrometry will be required to investigate the Q2 objects and verify if the above conclusions drawn for the Q1 sample also apply to the Q2 sample.




We conclude that the $s$-process rich and $s$-process non-enriched population of single Galactic post-AGB stars with reliable $Gaia$\,EDR3 astrometry covers the same metallicity range and luminosity distribution, revealing an intriguing chemical diversity observed amongst the sample of single post-AGB stars in the Galaxy. We could not find observationally derived parameters which distinguish the two groups, apart from \cfe\, and possibly \nfe. The intriguing  chemical diversity within a wide luminosity range shows that this is a fundamental shortcoming in our understanding of the chemical evolution of LIM stars. This suggests that the chemical evolution of LIM stars are not only dependent on the initial mass and metallicity but that additional parameters are needed to understand the diversity. Processes such as rotation \citep[e.g.]{denhartogh19}, overshoot physics \cite[e.g.][]{goriely18,kamath12}, extra mixing \citep[e.g.][]{karakas10,karakas14a,ventura15} and mass loss \citep[e.g.][]{cristallo11,karakas14a} can all affect the outcome of nucleosynthesis during and prior to the AGB phase. As the next step in our research we will model all objects individually to investigate the parameter space needed to explain the chemical diversity in the luminosity range we probe. In Paper II, we will therefore first investigate, in detail, the CNO abundances of the individual objects. Subsequently, we will model the full $s$-process abundance profiles to gain insight into which parameters in the stellar evolution modelling are most deterministic in the $s$-process nucleosynthesis and photospheric enrichment processes. 



\begin{acknowledgments}
DK  acknowledges  the  support  of  the  Australian  Research Council (ARC)  Discovery  Early  Career  Research  Award (DECRA) grant (DE190100813). This research was supported in part by the Australian Research Council Centre of Excellence for All Sky Astrophysics in 3 Dimensions (ASTRO 3D), through project number CE170100013. HVW acknowledges support from the Research Council of the KU Leuven under grant number C14/17/082. DK acknowledges the initial discussions she had with Prof. Peter Wood on the importance of $Gaia$ parallaxes for post-AGB stars and his continuous support. BJH acknowledges support from the US National Science Foundation (1413660). AIK was supported by the Australian Research Council Centre of Excellence for All Sky Astrophysics in 3 Dimensions (ASTRO 3D), through project number CE170100013. The authors thank the anonymous referee for their valuable suggestions and comments. 

\end{acknowledgments}

%

\vspace{5mm}
\facilities{$Gaia$\,EDR3}




 
\appendix

\section{Additional target details and the chemical abundances of the sample} 
\label{sec:chemabund}

Here we present additional information such as other names, the $Gaia$\,EDR3 identifier and the coordinates of the each of the individual objects. See Section~\ref{sec:target} for more details. We also present the relevant chemical abundances (taken from literature) for each of the individual objects. See Section~\ref{sec:DC} for more details. 

\begin{longrotatetable}
\begin{deluxetable*}{rrrrrrrrr}
\tabletypesize{\footnotesize}
\tablecolumns{9}
\tablewidth{0pt}
\small
\tabletypesize{\scriptsize}
\setlength{\tabcolsep}{3pt}
\tablecaption{A compilation of the various names for the target sample together with the $Gaia$ EDR3 identifiers, and the RA($^\circ$) and DEC($^\circ$). \label{table2}}
\tablehead{
\colhead{Index} & \colhead{IRAS Name} & \colhead{HR NAME} & \colhead{HD NAME} & \colhead{SAO Name} & \colhead{Other Names} & 
\colhead{$Gaia$ EDR3 Identifier} & \colhead{RA($^\circ$)} & \colhead{DEC($^\circ$)} \\
}
\startdata
\multicolumn{9}{c}{Post-AGB stars with $s$-process enrichment}\\ 
\hline
1	&	IRAS\,Z02229+6208	&	\nodata	&	\nodata	&	\nodata	&	\nodata	&	GAIA EDR3 513671461473684352	&	36.67411843	&	62.35611398	\\
2	&	IRAS\,04296+3429	&	\nodata	&	\nodata	&	\nodata	&	GLMP\,74	&	GAIA EDR3 173086700992466688	&	68.23740141	&	34.60344627	\\
3	&	IRAS\,05113+1347	&	\nodata	&	\nodata	&	\nodata	&	GLMP\,88	&	GAIA EDR3 3388902129107252992	&	78.53236845	&	13.84116291	\\
4	&	IRAS\,05341+0852	&	\nodata	&	\nodata	&	\nodata	&	GLMP\,106	&	GAIA EDR3 3334854780347915520	&	84.22938291	&	8.90240565	\\
5	&	IRAS\,06530-0213	&	\nodata	&	\nodata	&	\nodata	&	GLMP\,161, PN PM 1-24	&	GAIA EDR3 3105987960396950784	&	103.88258522	&	-2.29117496	\\
6	&	IRAS\,07134+1005	&	\nodata	&	HD\,56126	&	SAO\,96709	&	V$*$CY Cmi, BD+101470, GLMP\,174	&	GAIA EDR3 3156171118495247360	&	109.04274454	&	9.99665253	\\
7	&	IRAS\,07430+1115	&	\nodata	&	\nodata	&	\nodata	&	GLMP\,192	&	GAIA EDR3 3151417586128916864	&	116.46415377	&	11.13875852	\\
8	&	IRAS\,08143-4406	&	\nodata	&	\nodata	&	\nodata	&	GLMP\,206, PN PM 1-39	&	GAIA EDR3 5520238967817034880	&	124.01257847	&	-44.26794331	\\
9	&	IRAS\,08281-4850	&	\nodata	&	\nodata	&	\nodata	&	GLMP\,218, PN PM 1-40	&	GAIA EDR3 5515266327706463616	&	127.41895483	&	-49.00119212	\\
10	&	IRAS\,12360-5740	&	\nodata	&	\nodata	&	\nodata	&	GLMP\,334	&	GAIA EDR3 6060828565581083264	&	189.72126697	&	-57.94218386	\\
11	&	IRAS\,13245-5036	&	\nodata	&	\nodata	&	\nodata	&	GLMP\,347	&	GAIA EDR3 6070128028770373888	&	201.90386260	&	-50.86837655	\\
12	&	IRAS\,14325-6428	&	\nodata	&	\nodata	&	\nodata	&	\nodata	&	GAIA EDR3 5849962851220246016	&	219.14313660	&	-64.69197718	\\
13	&	IRAS\,14429-4539	&	\nodata	&	\nodata	&	\nodata	&	\nodata	&	GAIA EDR3 5906408788891928704	&	298.21959045	&	-17.03065000	\\
14	&	IRAS\,19500-1709	&	\nodata	&	HD\,187885	&	SAO\,163075	&	V$*$ V5112 Sgr, BD-175779, GLMP\,954	&	GAIA EDR3 6871175064823382912	&	300.49794561	&	32.79242099	\\
15	&	IRAS\,20000+3239	&	\nodata	&	\nodata	&	\nodata	&	GLMP\,963	&	GAIA EDR3 2034134414507432064	&	336.13092033	&	43.71970270	\\
16	&	IRAS\,22223+4327	&	\nodata	&	\nodata	&	\nodata	&	V$*$ V448 Lac, BD+424388, GLMP\,1058	&	GAIA EDR3 1958757291756223104	&	337.29322799	&	54.85174583	\\
17	&	IRAS\,22272+5435	&	\nodata	&	HD\,235858	&	SAO\,34504	&	V$*$ V354 Lac, BD+542787, GLMP\,1059	&	GAIA EDR3 2006425553228658816	&	353.18657229	&	62.06362764	\\
18	&	IRAS\,23304+6147	&	\nodata	&	\nodata	&	\nodata	&	GLMP\,1078	&	GAIA EDR3 2015785313459952128	&	221.55738007	&	-45.86812587	\\
\hline
\multicolumn{9}{c}{Post-AGB stars without $s$-process enrichment}\\
\hline
19	&	IRAS\,01259+6823	&	\nodata	&	\nodata	&	\nodata	&	\nodata	&	GAIA EDR3 532078488709487360	&	22.39034075	&	68.65402620	\\
20	&	IRAS\,08187-1905	&	\nodata	&	HD\,70379	&	\nodata	&	V$*$V552 Pup, BD-182290, GLMP\,209	&	GAIA EDR3 5707613169577769600	&	125.23791728	&	-19.25094348	\\
21	&	IRAS\,12175-5338	&	\nodata	&	\nodata	&	SAO\,239853	&	V$*$V1024 Cen, GLMP\,321	&	GAIA EDR3 6076326701687231872	&	185.06270566	&	-53.92538756	\\
22	&	\nodata	&	\nodata	&	HD\,107369	&	SAO\,203367	&	\nodata	&	GAIA EDR3 3469106382752903168	&	185.18720421	&	-32.55726017	\\
23	&	IRAS\,12538-2611	&	HR\,4912	&	HD\,112374	&	SAO\,181244	&	V$*$ LN Hya, 	&	GAIA EDR3 3497154104039422848	&	194.12560314	&	-26.46038652	\\
24	&	IRAS\,15039-4806	&	\nodata	&	HD\,133656	&	SAO\,225457	&	\nodata	&	GAIA EDR3 5903310335089068416	&	226.86432907	&	-48.29833370	\\
25	&	IRAS\,F16277-0724	&	HR\,6144	&	HD\,148743	&	SAO\,141206	&	\nodata	&	GAIA EDR3 4351018375858237952	&	247.62509302	&	-7.51444515	\\
26	&	IRAS\,17436+5003	&	\nodata	&	HD\,161796	&	SAO\,30548	&	V$*$V814 Her, BD+502457, GLMP\,639	&	GAIA EDR3 1367102319545324288	&	266.23111123	&	50.04424810	\\
27	&	IRAS\,18025-3906	&	\nodata	&	\nodata	&	\nodata	&	GLMP\,713	&	GAIA EDR3 4035907203854415488	&	271.51374773	&	-39.09910702	\\
28	&	IRAS\,18095+2704	&	\nodata	&	HD\,335675	&	\nodata	&	V$*$ V887 Her, GLMP\,735	&	GAIA EDR3 4580154606223711872	&	272.87773398	&	27.08767172	\\
29	&	IRAS\,19386+0155	&	\nodata	&	\nodata	&	\nodata	&	V$*$V1648 Aql	&	GAIA EDR3 4240112390324832384	&	295.28454494	&	2.04197942	\\
30	&	IRAS\,19475+3119	&	\nodata	&	HD\,331319	&	\nodata	&	V$*$V2513 Cyg, BD+313797, GLMP\,951	&	GAIA EDR3 2033763428091006720	&	297.37317348	&	31.45450606	\\
31	&	IRAS\,20023-1144	&	HR\,7671	&	HD\,190390	&	 SAO\,163245	&	V$*$V1401 Aql, BD-125641	&	GAIA EDR3 4190636669164572928	&	301.27254423	&	-11.59948519	\\
\enddata
\end{deluxetable*}
\end{longrotatetable}

\begin{longrotatetable}
\begin{deluxetable*}{rrrrrrrrrrrrrr}
\tabletypesize{\footnotesize}
\tablecolumns{14}
\tablewidth{0pt}
\small
\tabletypesize{\scriptsize}
\setlength{\tabcolsep}{3pt}
\tablecaption{Chemical abundances of the $s$–process enriched and non $s$–process enriched single Galactic post–AGB stars. \label{table3}}
\tablehead{
\colhead{Index} & \colhead{Object Name} & \colhead{\feh} & \colhead{\co} & \colhead{\ofe} & \colhead{\cfe} & 
\colhead{\nfe} & \colhead{\zrfe} & \colhead{\sfe} & \colhead{\lsfe} & \colhead{\hsfe} & \colhead{\hsls} & \colhead{Ref.}
} 
\startdata
\multicolumn{13}{c}{Post–AGB stars with $s$–process enrichment}\\ 
\hline
1	&	IRAS\,Z02229+6208	&	–0.45 ± 0.14	&	\nodata	&	\nodata	&	0.78 ± 0.15	&	1.19 ± 0.30	&	2.22 ± 0.13	&	1.4 ± 0.15	&	2.03 ± 0.12	&	1.12 ± 0.03	&	–0.91 ± 0.12	&	1	\\
2	&	IRAS\,04296+3429	&	–0.62 ± 0.11	&	\nodata	&	\nodata	&	0.8 ± 0.2	&	0.39 ± 0.01	&	1.34 ± 0.23	&	1.5 ± 0.23	&	1.7 ± 0.23	&	1.5 ± 0.17	&	–0.2 ± 0.23	&	2	\\
3	&	IRAS\,05113+1347	&	–0.49 ± 0.15	&	2.42 ± 0.40	&	0.01 ± 0.27	&	0.65 ± 0.16	&	\nodata	&	1.36 ± 0.15	&	1.54 ± 0.07	&	1.33 ± 0.13	&	1.65 ± 0.07	&	0.32 ± 0.15	&	3	\\
4	&	IRAS\,05341+0852	&	–0.54 ± 0.11	&	1.06 ± 0.30	&	0.75 ± 0.11	&	1.03 ± 0.10	&	\nodata	&	1.76 ± 0.10	&	2.12 ± 0.05	&	1.87 ± 0.08	&	2.24 ± 0.06	&	0.37 ± 0.10	&	3	\\
5	&	IRAS\,06530–0213	&	–0.32 ± 0.11	&	1.66 ± 0.39	&	0.35 ± 0.11	&	0.83 ± 0.13	&	\nodata	&	1.60 ± 0.10	&	1.94 ± 0.06	&	1.75 ± 0.09	&	2.04 ± 0.08	&	0.29 ± 0.13	&	3	\\
6	&	IRAS\,07134+1005	&	–0.91 ± 0.20	&	1.24 ± 0.29	&	0.81 ± 0.19	&	1.16 ± 0.22	&	0.57 ± 0.19	&	1.61 ± 0.09	&	1.63 ± 0.14	&	1.64 ± 0.13	&	1.63 ± 0.20	&	–0.01 ± 0.24	&	3	\\
7	&	IRAS\,07430+1115	&	–0.31 ± 0.15	&	1.71 ± 0.30	&	0.30  ± 0.22	&	0.79 ± 0.13	&	\nodata	&	1.22 ± 0.15	&	1.47 ± 0.06	&	1.30 ± 0.14	&	1.55 ± 0.06	&	0.25 ± 0.15	&	3	\\
8	&	IRAS\,08143–4406	&	–0.43 ± 0.11	&	1.66 ± 0.39	&	0.19 ± 0.13	&	0.71 ± 0.10	&	0.01 ± 0.22	&	1.63 ± 0.11	&	1.65 ± 0.05	&	1.77 ± 0.08	&	1.58 ± 0.06	&	–0.19 ± 0.11	&	3	\\
9	&	IRAS\,08281–4850	&	–0.26 ± 0.11	&	2.34 ± 0.42	&	0.12 ± 0.11	&	0.75 ± 0.21	&	\nodata	&	1.42 ± 0.11	&	1.58 ± 0.09	&	1.57 ± 0.11	&	1.58 ± 0.12	&	0.01 ± 0.17	&	3	\\
10	&	IRAS\,12360–5740	&	–0.40 ± 0.15	&	0.45 ± 0.20	&	0.31 ± 0.05	&	0.27 ± 0.18	&	0.22 ± 0.32	&	1.70 ± 0.17	&	1.88 ± 0.20	&	1.73 ± 0.20	&	2.02 ± 0.20	&	0.29 ± 0.20	&	4	\\
11	&	IRAS\,13245–5036	&	–0.30 ± 0.10	&	1.11 ± 0.30	&	0.26 ± 0.13	&	0.57 ± 0.21	&	\nodata	&	1.72 ± 0.15	&	1.88 ± 0.09	&	1.56 ± 0.14	&	2.03 ± 0.11	&	0.47 ± 0.18	&	3	\\
12	&	IRAS\,14325–6428	&	–0.56 ± 0.10	&	2.27 ± 0.40	&	0.57 ± 0.09	&	1.18 ± 0.23	&	0.18 ± 0.20	&	1.16 ± 0.16	&	1.30 ± 0.14	&	1.25 ± 0.15	&	1.33 ± 0.19	&	0.08 ± 0.24	&	3	\\
13	&	IRAS\,14429–4539	&	–0.18 ± 0.11	&	1.29 ± 0.26	&	0.31 ± 0.12	&	0.68 ± 0.23	&	\nodata	&	1.46 ± 0.17	&	1.41 ± 0.08	&	1.29 ± 0.15	&	1.47 ± 0.10	&	0.18 ± 0.08	&	3	\\
14	&	IRAS\,19500–1709	&	–0.59 ± 0.10	&	1.02 ± 0.17	&	0.72 ± 0.04	&	0.99 ± 0.06	&	0.41 ± 0.30	&	1.34 ± 0.10	&	1.35 ± 0.21	&	1.37 ± 0.29	&	1.34 ± 0.30	&	–0.03 ± 0.41	&	3	\\
15	&	IRAS\,20000+3239	&	–1.4 ± 0.2	&	\nodata	&	\nodata	&	1.7 ± 0.2	&	2.1 ± 0.2	&	1 ± 0.2	&	1.4 ± 0.2	&	1.1 ± 0.2	&	1.47 ± 0.10	&	0.34 ± 0.2	&	5	\\
16	&	IRAS\,22223+4327	&	–0.30 ± 0.11	&	1.04 ± 0.22	&	0.31 ± 0.06	&	0.59 ± 0.06	&	0.15 ± 0.30	&	1.35 ± 0.06	&	1.03 ± 0.05	&	1.34 ± 0.07	&	0.88 ± 0.07	&	–0.46 ± 0.10	&	3	\\
17	&	IRAS\,22272+5435	&	–0.77 ± 0.12	&	1.46 ± 0.26	&	0.63 ± 0.02	&	1.05 ± 0.07	&	\nodata	&	1.54 ± 0.08	&	1.80 ± 0.05	&	1.61 ± 0.08	&	1.90 ± 0.07	&	0.28 ± 0.11	&	3	\\
18	&	IRAS\,23304+6147	&	–0.81 ± 0.2	&	2.8 ± 0.2	&	0.17 ± 0.03	&	0.91 ± 0.12	&	0.47 ± 0.15	&	1.26 ± 0.23	&	1.60 ± 0.25	&	1.55 ± 0.23	&	1.63 ± 0.21	&	0.09 ± 0.24	&	6	\\
\hline
\multicolumn{13}{c}{Post–AGB stars without $s$–process enrichment}\\
\hline
19	&	IRAS\,01259+6823	&	–0.60 ± 0.1	&	0.4 ± 0.3	&	0.31 ± 0.06	&	0.18 ± 0.3	&	\nodata	&	0.12 ± 0.1	&	0.3 ± 0.1	&	\nodata	&	\nodata	&	\nodata	&	7	\\
20	&	IRAS\,08187–1905	&	–0.60 ± 0.1	&	\nodata	&	0.26 ± 0.1	&	0.62 ± 0.3	&	0.49 ± 0.3	&	0.25 ± 0.1	&	\nodata	&	\nodata	&	\nodata	&	\nodata	&	7	\\
21	&	SAO 239853	&	–0.81 ± 0.1	&	\nodata	&	0.8 ± 0.2	&	0.4 ± 0.2	&	0.6 ± 0.2	&	\nodata	&	–0.4 ± 0.2	&	\nodata	&	\nodata	&	\nodata	&	8	\\
22	&	HD\,107369	&	–1.1 ± 0.1	&	\nodata	&	0 ± 0.2	&	$<$–0.2	&	0.49 ± 0.3	&	\nodata	&	–0.1 ± 0.2	&	\nodata	&	\nodata	&	\nodata	&	8	\\
23	&	HD\, 112374	&	–1.2 ± 0.1	&	\nodata	&	0.8 ± 0.2	&	0.1 ± 0.2	&	0.5 ± 0.2	&	\nodata	&	–0.3 ± 0.2	&	\nodata	&	\nodata	&	\nodata	&	8	\\
24	&	HD\,133656	&	–0.7 ± 0.1	&	\nodata	&	0.6 ± 0.2	&	0.3 ± 0.2	&	0.5 ± 0.2	&	\nodata	&	–0.4 ± 0.2	&	\nodata	&	\nodata	&	\nodata	&	8	\\
25	&	HR\, 6144	&	–0.4  ± 0.1	&	\nodata	&	0.3 ± 0.2	&	0.3 ± 0.2	&	0.9 ± 0.2	&	\nodata	&	0.2 ± 0.2	&	\nodata	&	\nodata	&	\nodata	&	8	\\
26	&	HD\,161796	&	–0.3 ± 0.1	&	\nodata	&	0.4 ± 0.2	&	0.3 ± 0.2	&	1.1 ± 0.2	&	\nodata	&	0 ± 0.2	&	\nodata	&	\nodata	&	\nodata	&	8	\\
27	&	IRAS\,18025–3906	&	–0.51 ± 0.15	&	0.43	&	0.56 ± 0.2	&	0.46 ± 0.2	&	0.74 ± 0.2	&	–0.84 ± 0.04	&	\nodata	&	\nodata	&	\nodata	&	\nodata	&	9	\\
28	&	HD\,335675	&	–0.9 ± 0.2	&	0.25	&	0.77 – 0.19	&	0.4 – 0.35	&	$<$0.27 	&	–0.36 ± 0.1	&	\nodata	&	\nodata	&	\nodata	&	\nodata	&	10	\\
29	&	IRAS\,19386+0155	&	–1.1 ± 0.14	&	\nodata	&	\nodata	&	0.1 ± 0.2	&	\nodata	&	\nodata	&	–0.3 ± 0.2	&	\nodata	&	\nodata	&	\nodata	&	11	\\
30	&	IRAS\,19475+3119	&	–0.24 ± 0.15	&	0.19	&	0.30 ± 0.02	&	–0.09 ± 0.30	&	\nodata	&	\nodata	&	–0.30 ± 0.1	&	\nodata	&	\nodata	&	\nodata	&	12	\\
31	&	HR\,7671	&	–1.6 ± 0.1	&	0.05	&	0.46 ± 0.05	&	–0.57 ± 0.13	&	0.51 ± 0.16	&	0.44 ± 0.15	&	\nodata	&	\nodata	&	\nodata	&	\nodata	&	13	\\
\enddata
\tablenotetext{*}Note: The index \sfe\, is the mean of the relative abundances of the elements for the 'ls' and 'hs' indices. Typically, the 'ls' index uses the mean of the relative abundances of Y and Zr and the 'hs' index uses the mean of the relative abundances of La, Ce, Nd and Sm. \hsls\,=\,\hsfe\,$–$\,\lsfe. More details on the derived abundances and abundance ratios can be found in the individual studies mentioned in column 'Ref'. The column 'Ref.' indicates the individual chemical abundance study: $^{1}$\cite{reddy99}, $^{2}$\cite{vanwinckel00}, $^{3}$\cite{desmedt16}, $^{4}$\cite{pereira11}, $^{5}$\cite{klochkova06}, $^{6}$\cite{ReyniersThesis}, $^{7}$\cite{rao12}, $^{8}$\cite{vanwinckel97a}, $^{9}$\cite{molina19}, $^{10}$\cite{sahin11}, $^{11}$\cite{pereira04}, $^{12}$\cite{arellanoferro01}, $^{13}$\cite{reyniers05}.
\end{deluxetable*}
\end{longrotatetable}

\section{Spectral Energy Distributions (SEDs)} 
\label{sec:SEDs}

In this appendix we show the SEDs of the target sample of post-AGB stars (Q1 and Q2 objects). See Section~\ref{sec:lum+mass} for full details on the SED fitting. With regard to the photometric data points, as mentioned in Section~\ref{sec:lum+mass}, we queried the photometry from the Vizier database \citep{ochsenbein00}. We refer to Appendix A of \citet{oomen18} for the individual photometric catalogues used.

\begin{figure*}[h!]
{\gridline{\fig{IRAS_Z02229+6208SED.jpg}{0.35\textwidth}{1) IRASZ02229+6208}
          \fig{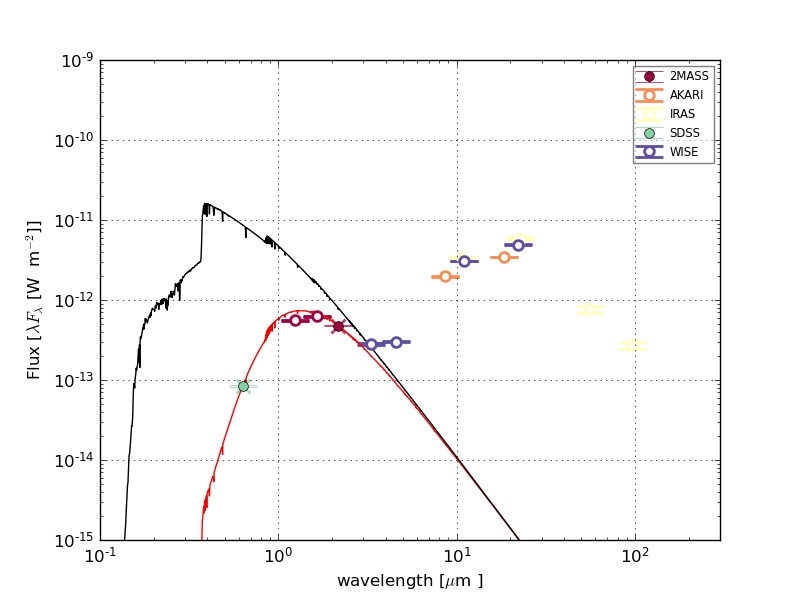}{0.35\textwidth}{2) IRAS04296+3429}
          \fig{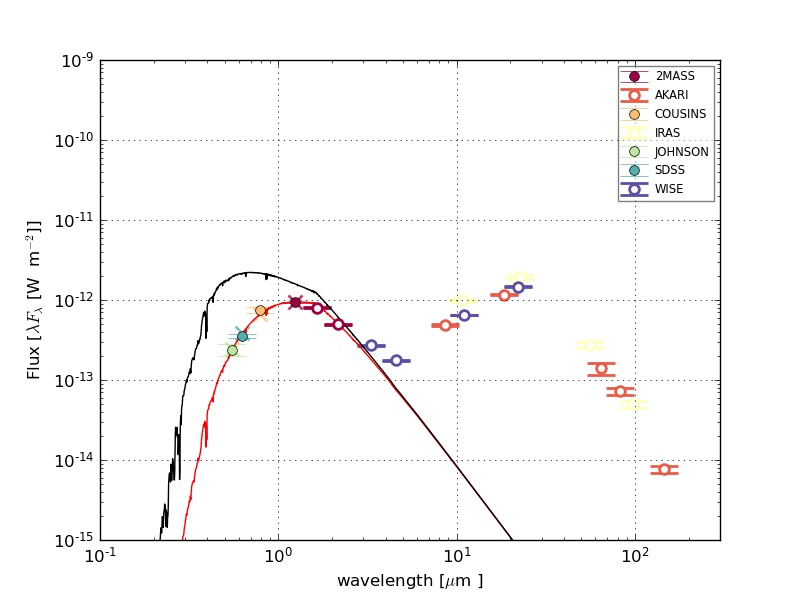}{0.35\textwidth}{3) IRAS05113+1347}}
\gridline{\fig{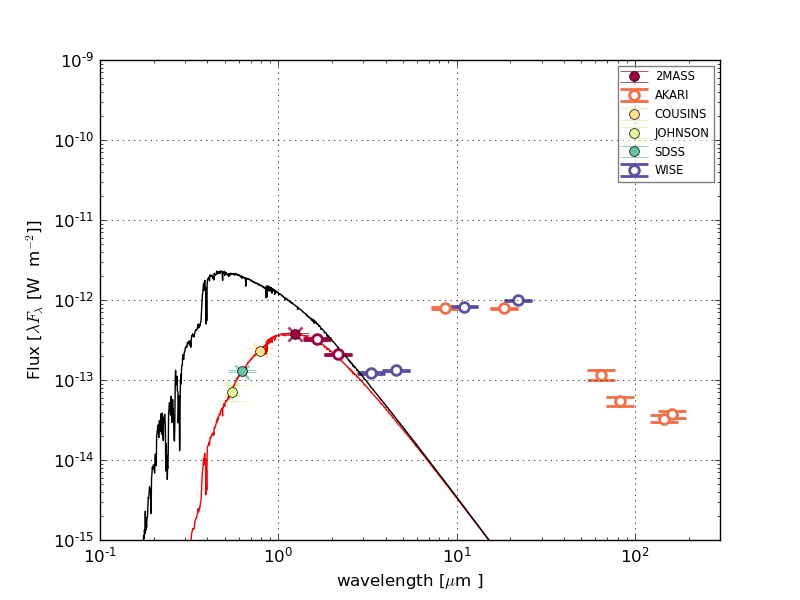}{0.35\textwidth}{4) IRAS05341+0852}
          \fig{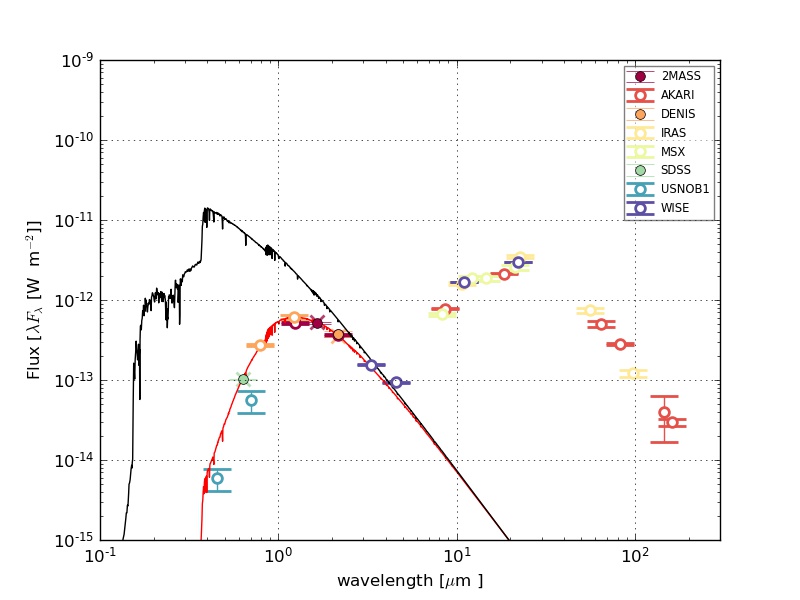}{0.35\textwidth}{5) IRAS06530-0213}
          \fig{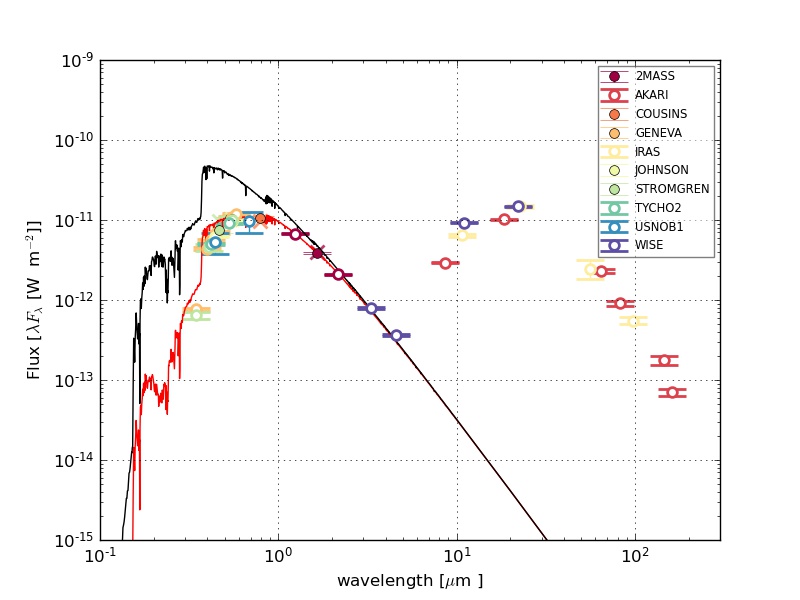}{0.35\textwidth}{6) IRAS07134+1005}}
\gridline{\fig{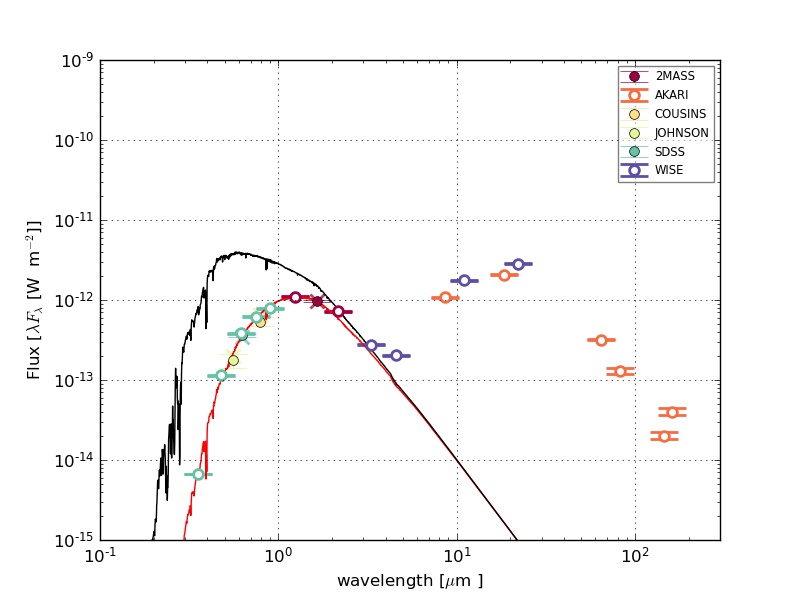}{0.35\textwidth}{7) IRAS07430+1115}
          \fig{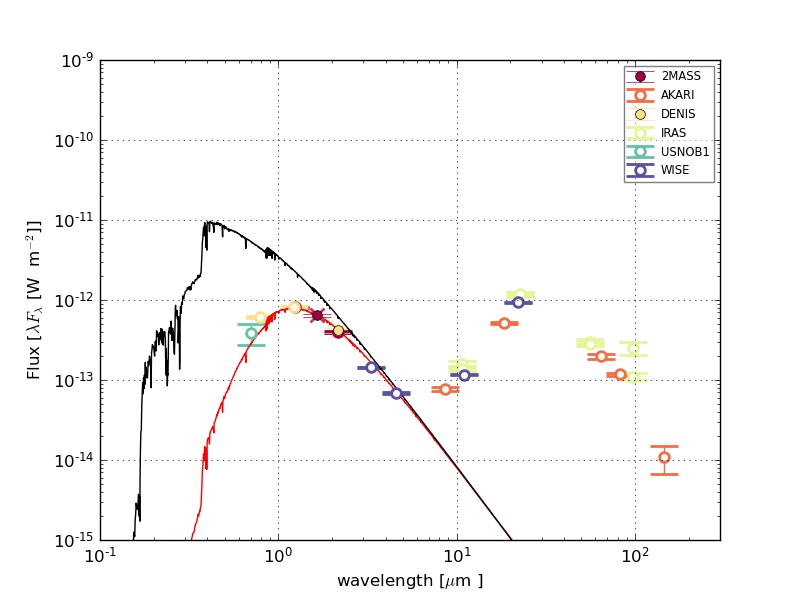}{0.35\textwidth}{8) IRAS08143-4406}
          \fig{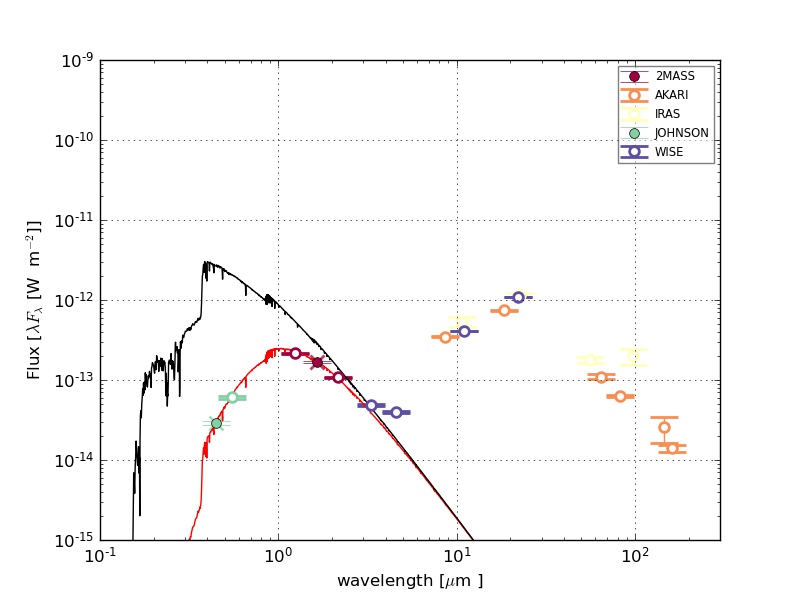}{0.35\textwidth}{9) IRAS08281-4850}}}
\caption{List of all SEDs of the post-AGB stars in our target sample. The order of the objects is same as that in Table~\ref{table1}. The black curve represents the atmospheric model. Details of the atmosphere model (e.g., \Teff, and E($B$\,$-$\,$V$) are listed in Table~\ref{table1}. The red solid curve is the reddened atmospheric model (see Section~\ref{sec:lum+mass} for more details. The various symbols correspond to the observed photometric data points where different colours denote different surveys mentioned on the plot.}
\label{fig:s_sed}
\end{figure*}
\begin{figure*}
\centering
{\gridline{\fig{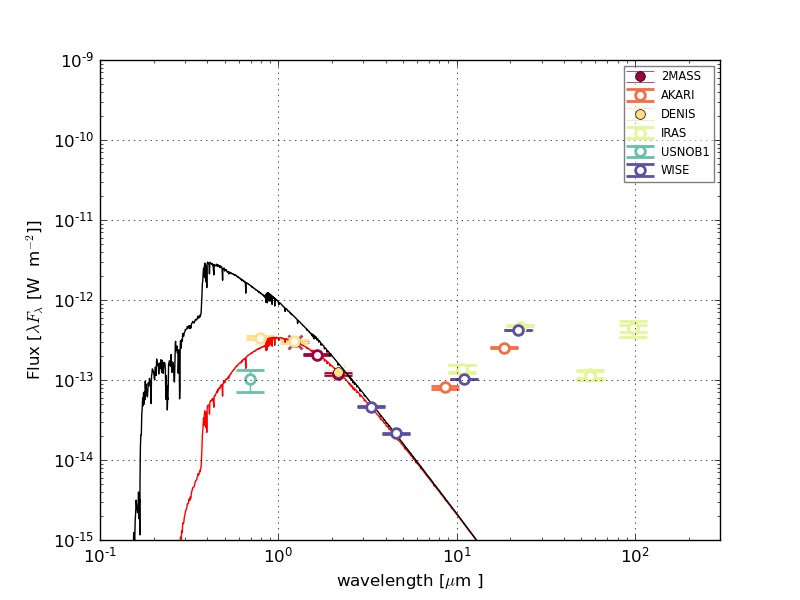}{0.35\textwidth}{10) IRAS12360-5740}
          \fig{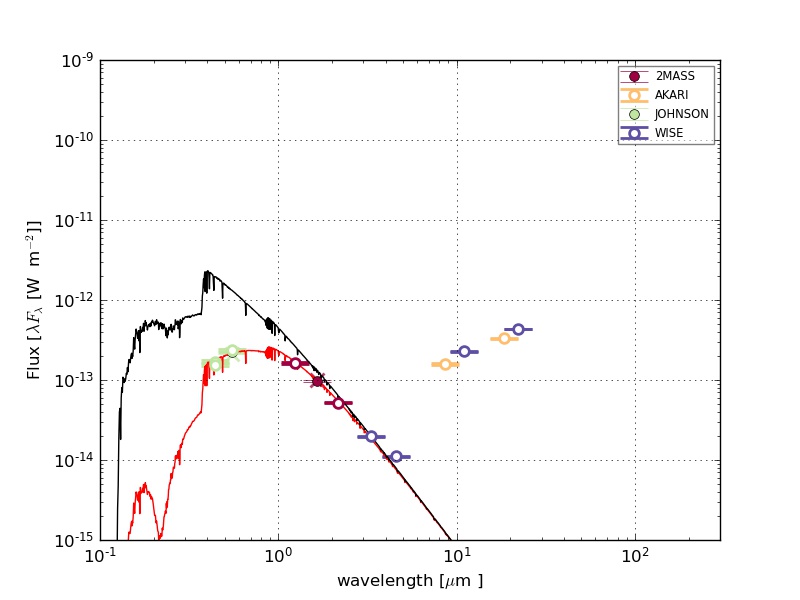}{0.35\textwidth}{11) IRAS13245-5036}
          \fig{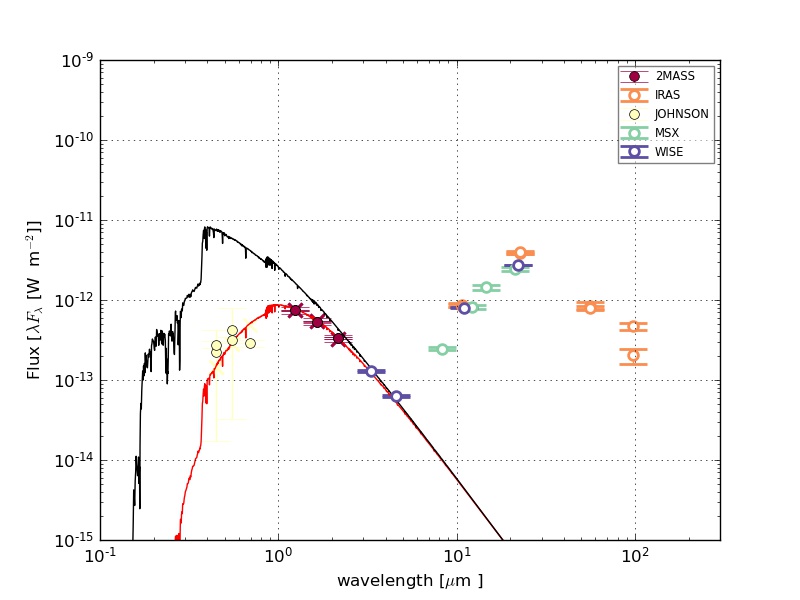}{0.35\textwidth}{12) IRAS14325-6428}}
\gridline{\fig{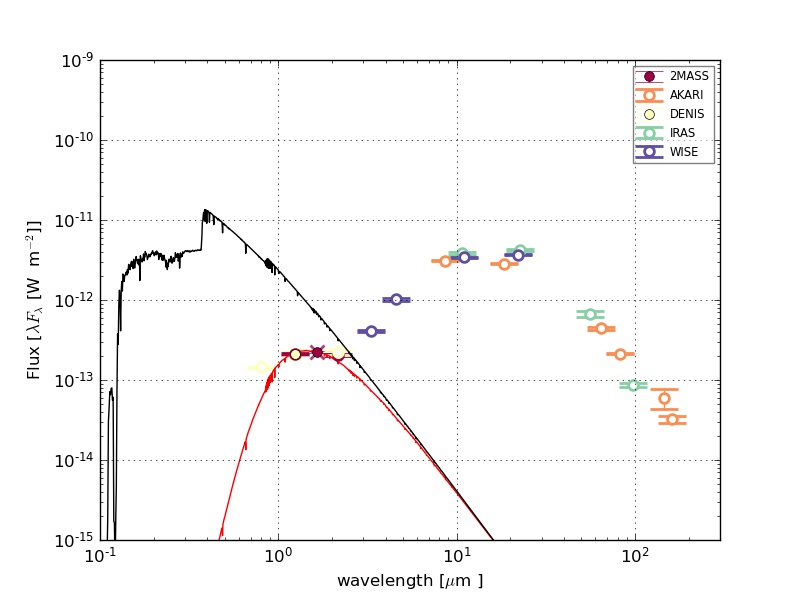}{0.35\textwidth}{13) IRAS14429-4539}
          \fig{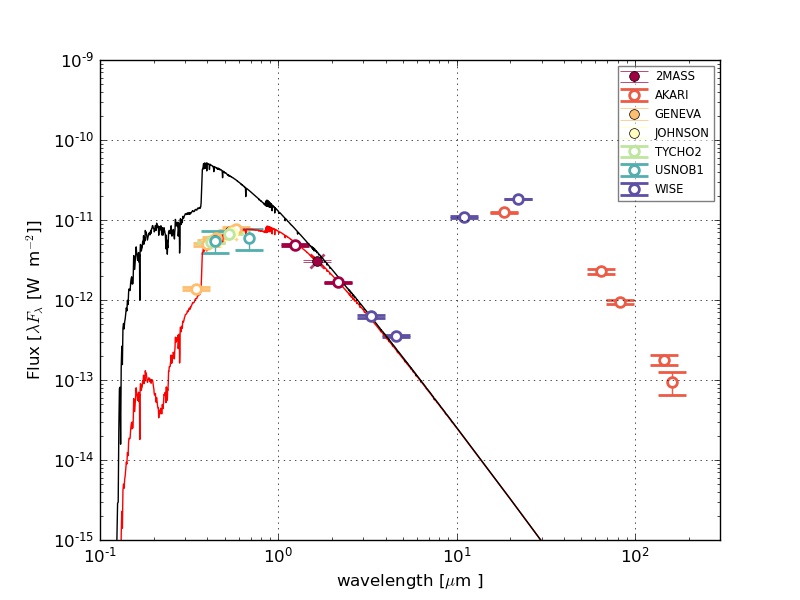}{0.35\textwidth}{14) IRAS19500-1709}
          \fig{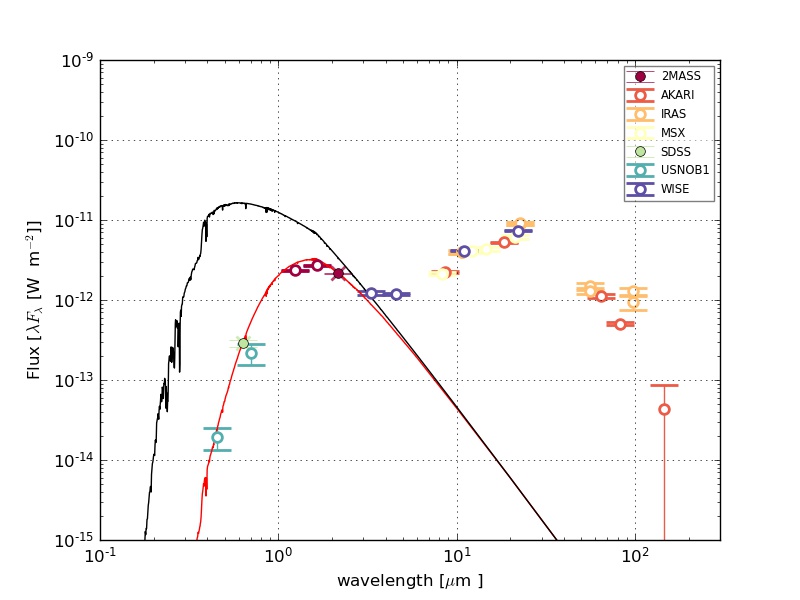}{0.35\textwidth}{15) IRAS20000+3239}}
\gridline{\fig{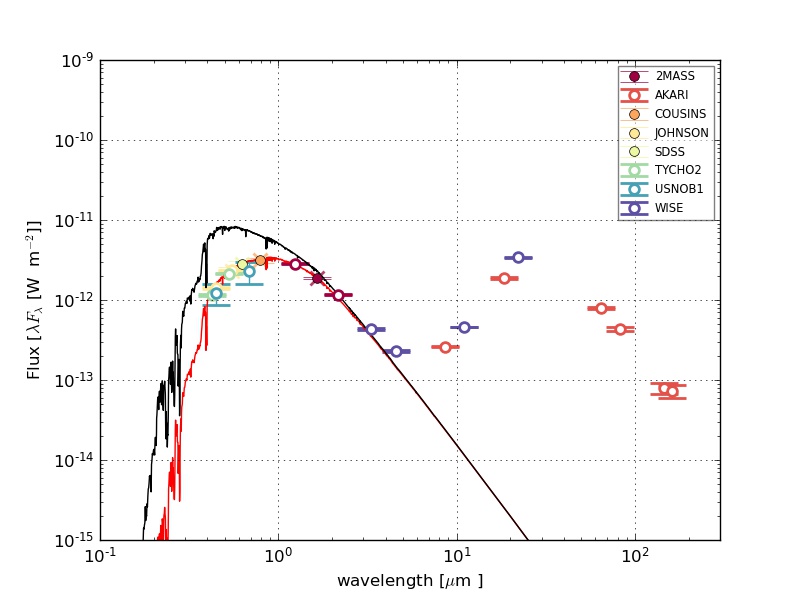}{0.35\textwidth}{16) IRAS22223+4327}
          \fig{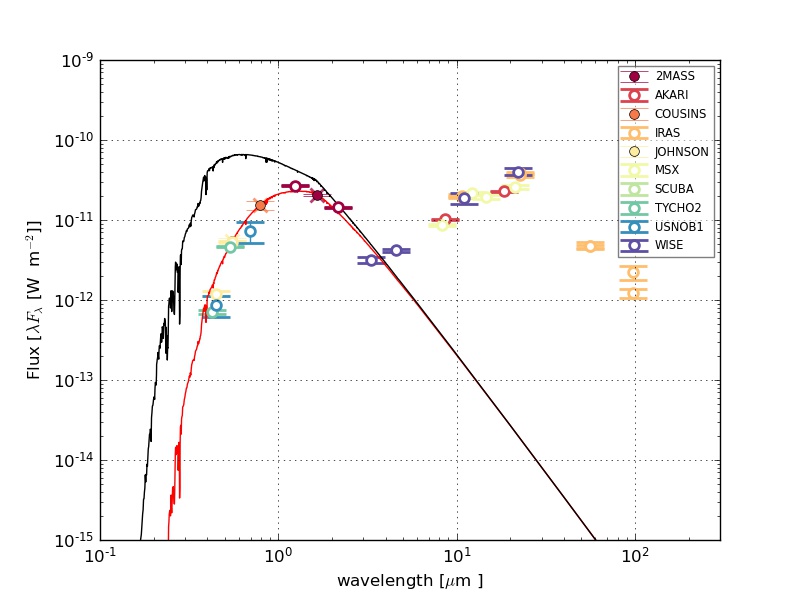}{0.35\textwidth}{17) IRAS22272+5435}
          \fig{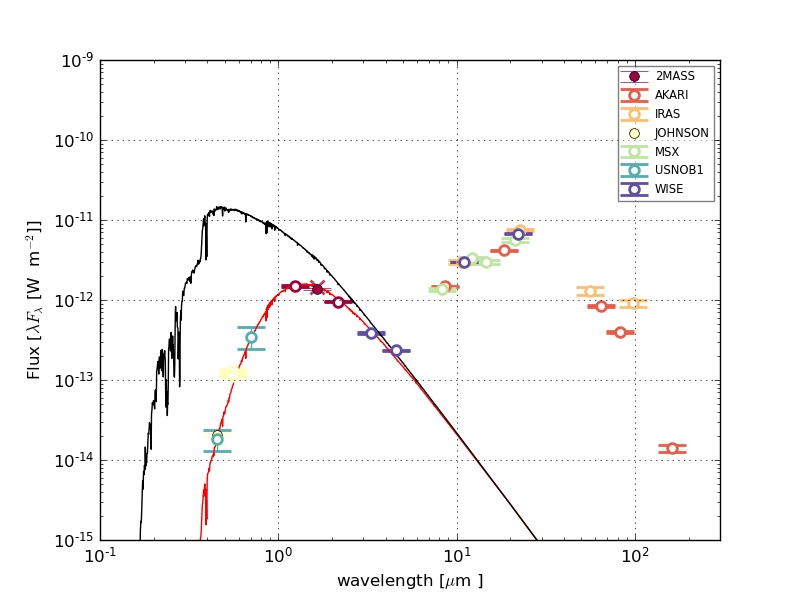}{0.35\textwidth}{18) IRAS23304+6147}}}
\gridline{\fig{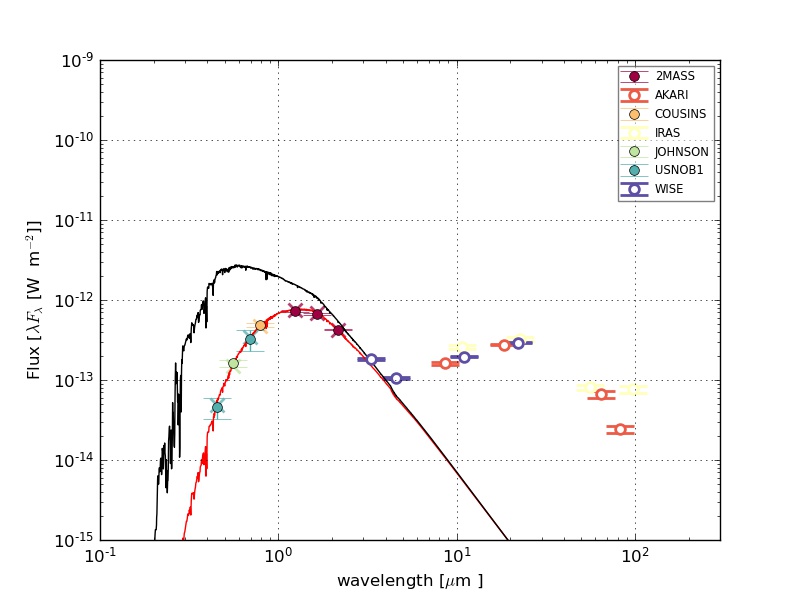}{0.35\textwidth}{19) IRAS01259+6823}
          \fig{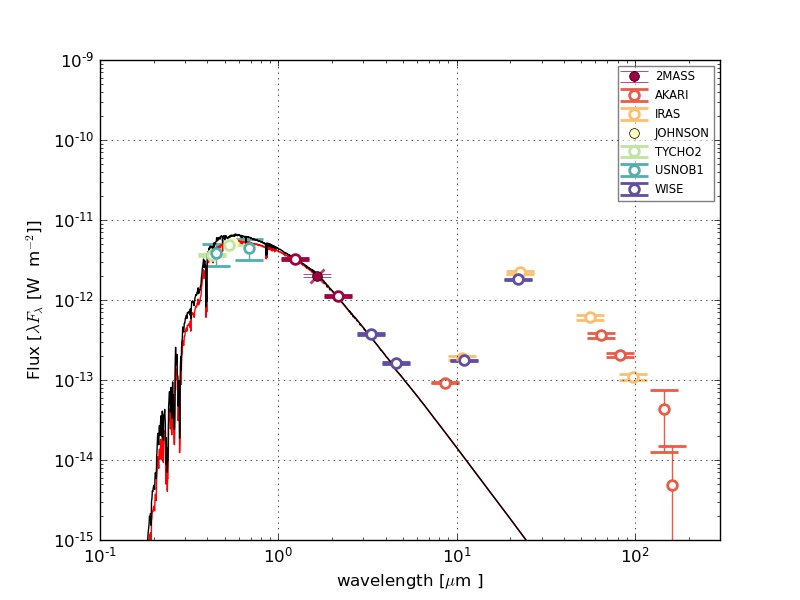}{0.35\textwidth}{20) IRAS08187-1905}
          \fig{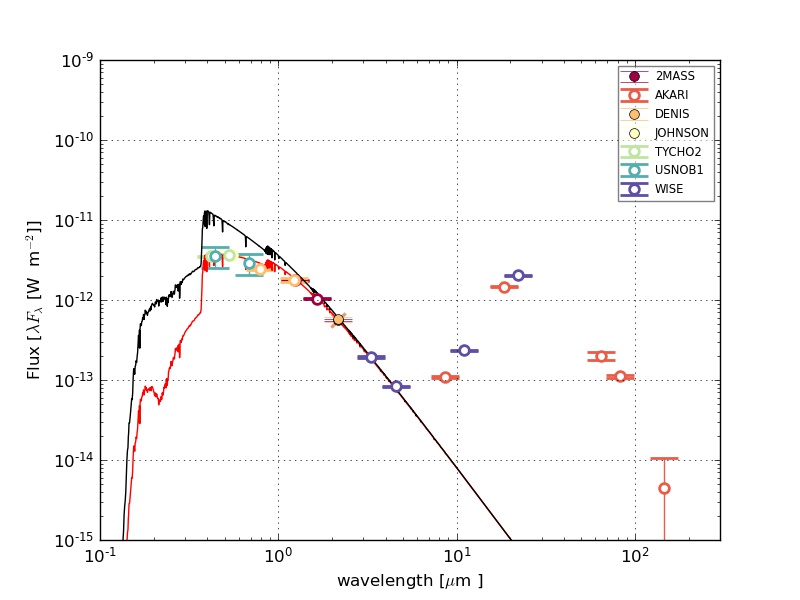}{0.35\textwidth}{21) SAO239853}}
\caption{Figure~\ref{fig:s_sed} continued.}
\end{figure*}
\begin{figure*}
\centering
\gridline{\fig{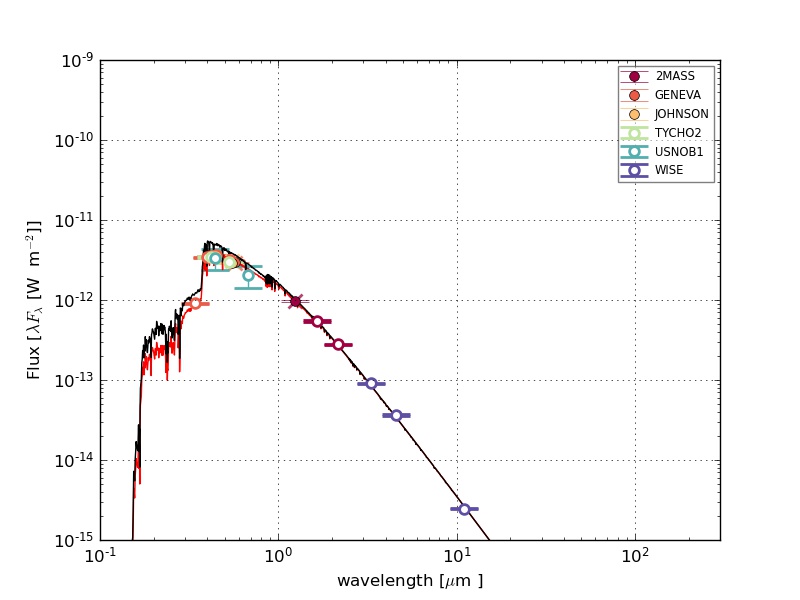}{0.35\textwidth}{22) HD107369}
          \fig{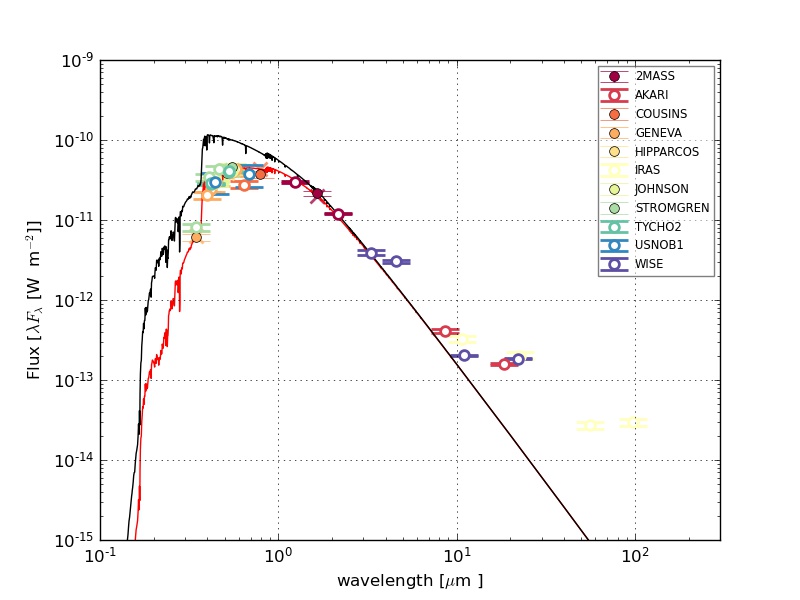}{0.35\textwidth}{23) HD112374}
          \fig{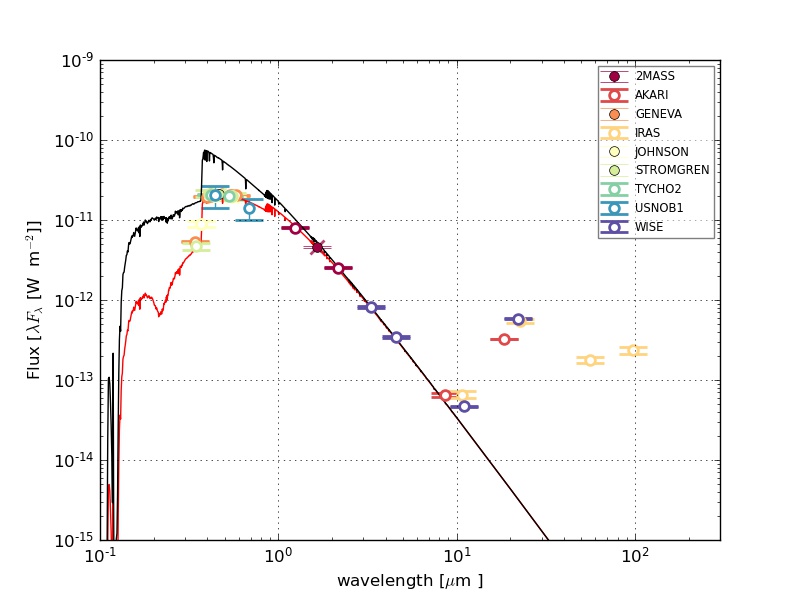}{0.35\textwidth}{24) HD133656}}
\gridline{\fig{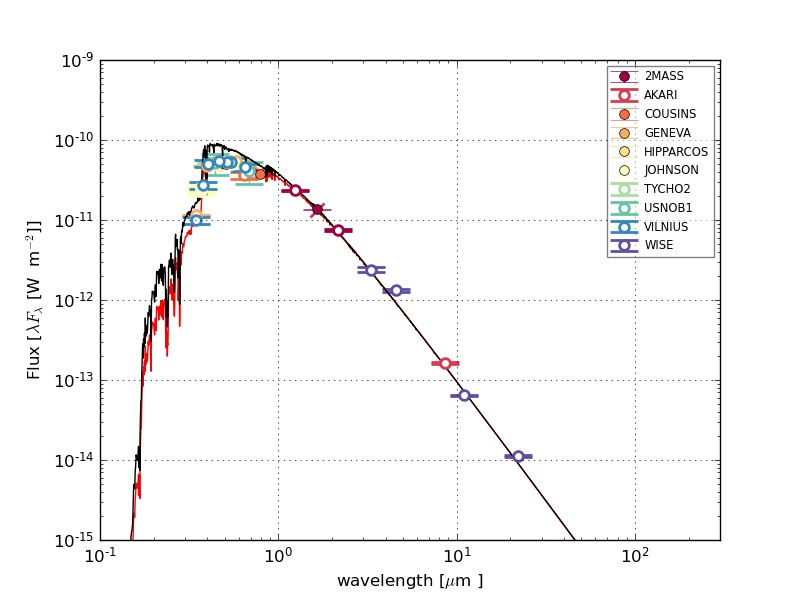}{0.35\textwidth}{25) HR6144}
          \fig{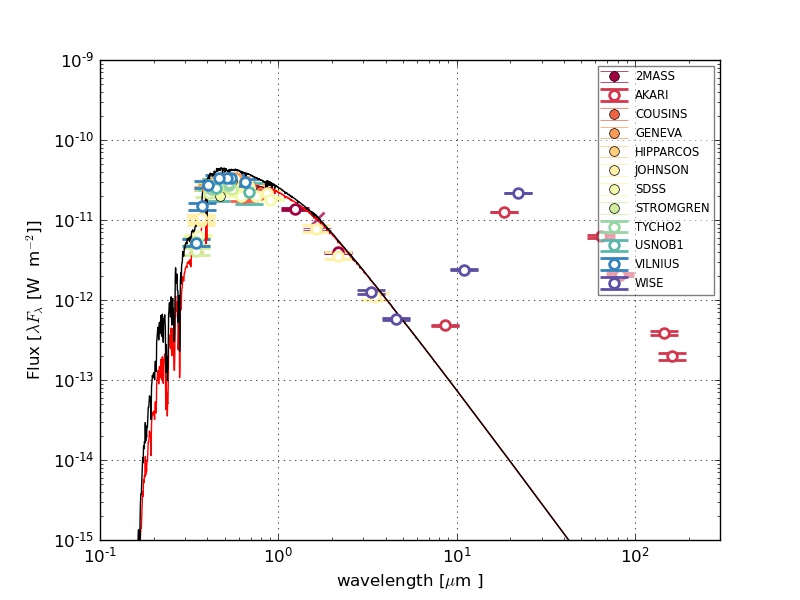}{0.35\textwidth}{26) HD161796}
          \fig{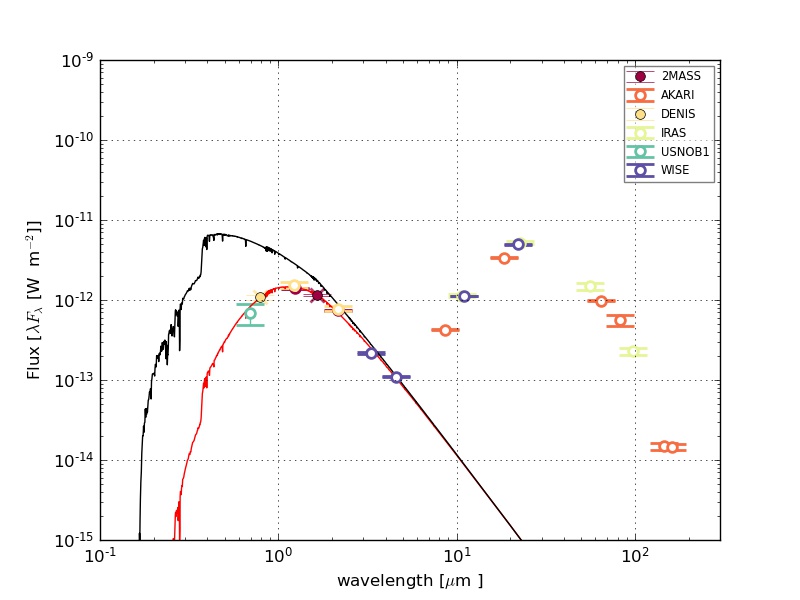}{0.35\textwidth}{27) IRAS18025-3906}}
\gridline{\fig{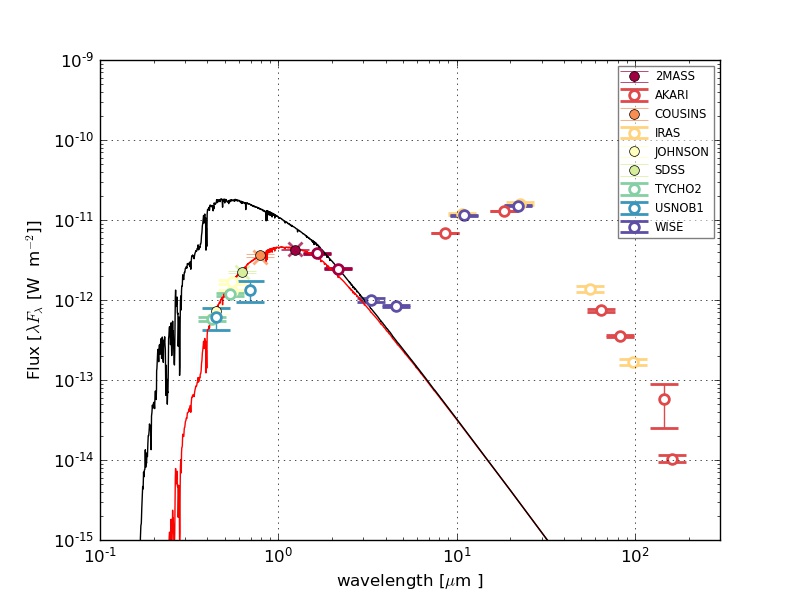}{0.35\textwidth}{28) IRAS18095+2704}
          \fig{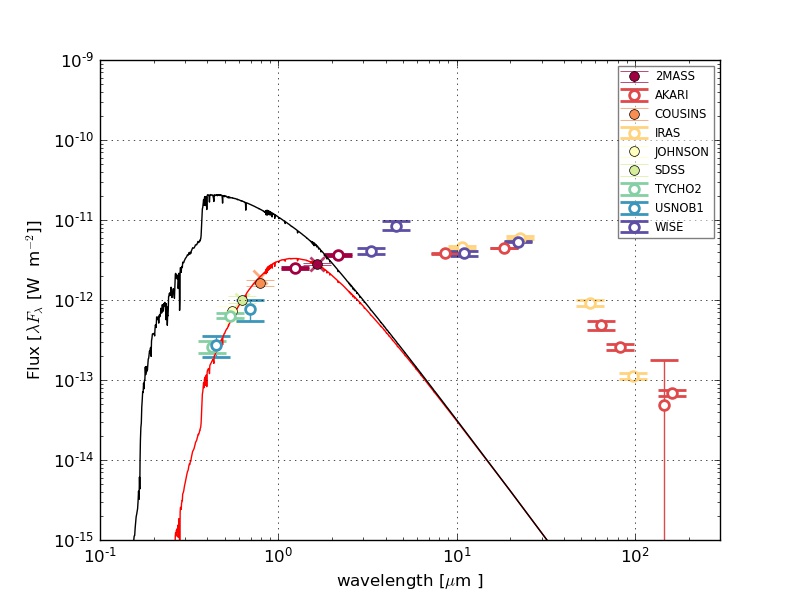}{0.35\textwidth}{29) IRAS19386+0155}
          \fig{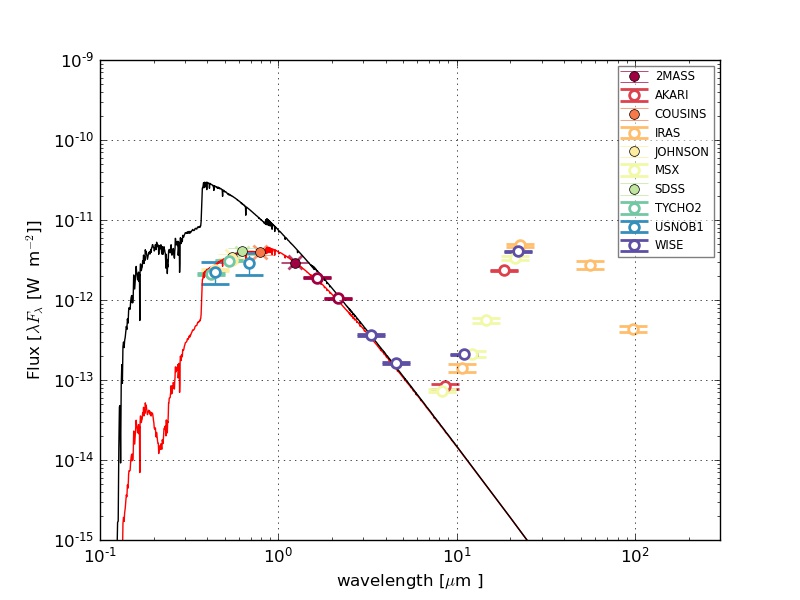}{0.35\textwidth}{30) IRAS19475+3119}}
\gridline{\fig{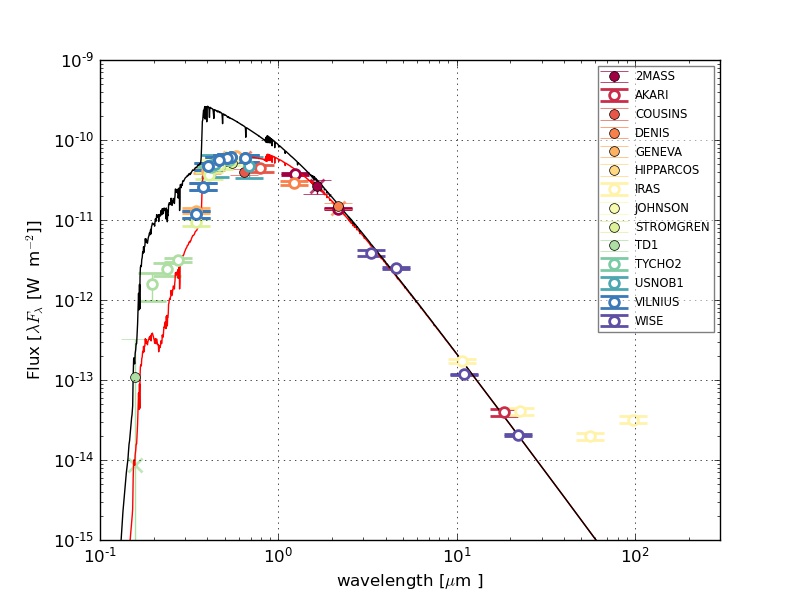}{0.35\textwidth}{31) HR7671}}
\caption{Figure~\ref{fig:s_sed} continued.}
\end{figure*}

\section{Positions of the Q1 and Q2 Galactic post-AGB single stars in the HR-diagram} 
\label{sec:hr_p1_p2}

\begin{figure*}
\gridline{\fig{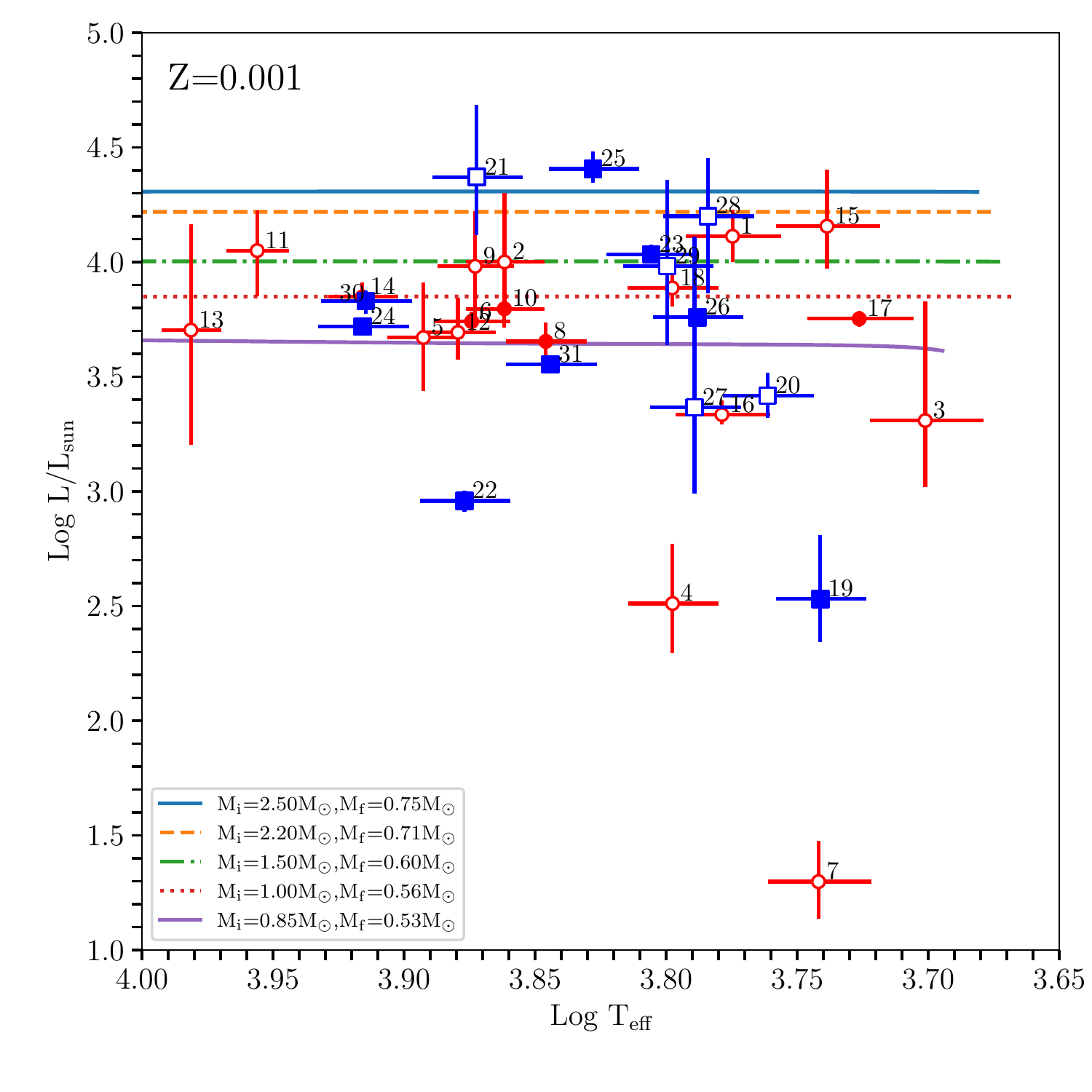}{0.5\textwidth}{}
          \fig{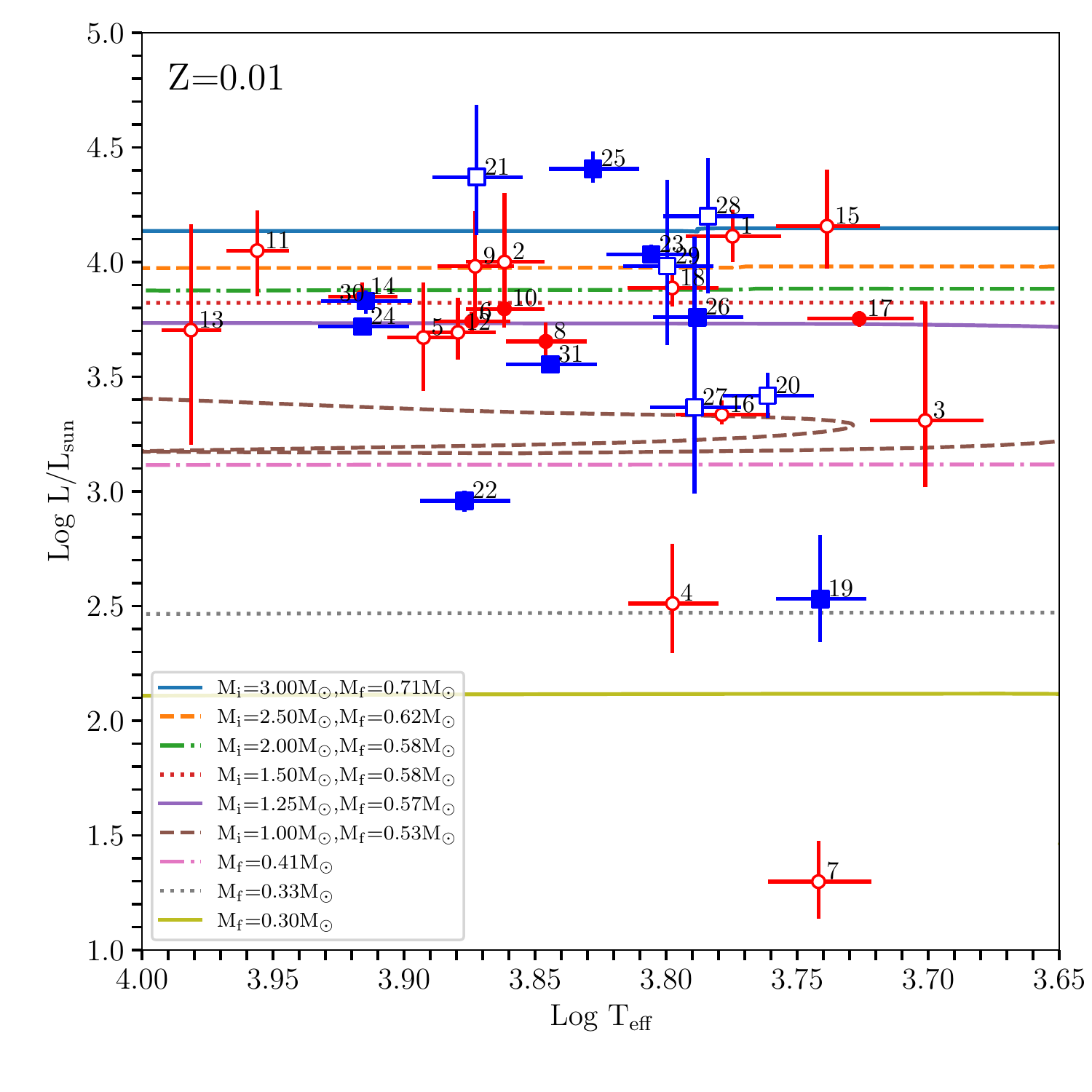}{0.5\textwidth}{}
          }
\caption{Positions of the Q1 and Q2 post-AGB stars in the HR diagram. The red filled-circles represent the $s$-process enriched Q1 objects and the blue filled-squares represent the non $s$-process enriched Q1 objects. The red open-circles represent the $s$-process enriched Q2 objects and the blue open-squares represent the non $s$-process enriched Q2 objects. The numbers represent the individual object numbers as listed in Table~\ref{table1}. Also shown are the available post-AGB evolutionary tracks of \citet{bertolami16} with $Z$\,=\,0.001 (right panel) and $Z$\,=\,0.01 (left panel). See text for more details. \label{fig:hr_p1_p2}}
\end{figure*}

\bibliography{mnemonic,devlib2021}
\bibliographystyle{aasjournal}



\end{document}